\documentstyle[12pt,aaspp4,epsf]{article}
\def\fun#1#2{\lower3.6pt\vbox{\baselineskip0pt\lineskip.9pt
  \ialign{$\mathsurround=0pt#1\hfil##\hfil$\crcr#2\crcr\sim\crcr}}}
\def\lap{\mathrel{\mathpalette\fun <}}
\def\gap{\mathrel{\mathpalette\fun >}}

\def\etal{{\it et al.}}

\def\mass{{\cal M}}

\def\Msolar{{\mass_\odot}}

\slugcomment{To Appear in {\bf The Astrophysical Journal}, Vol. {\bf 471}, 1996}

\lefthead{Merritt \& Valluri}
\righthead{Chaos and Mixing in Galaxies}

\begin{document}

\title{Chaos and Mixing in Triaxial Stellar Systems}

\author{David Merritt and Monica Valluri\footnote[1]{Current address: 
Department of Astronomy, 
Columbia University, New York, NY 10027}}
\affil{Department of Physics and Astronomy, Rutgers University,
New Brunswick, NJ 08855}

\begin{abstract}
We investigate the timescales for stochasticity and chaotic mixing 
in a family of triaxial potentials that mimic the distribution of 
light in elliptical galaxies.
Some of the models include central point masses designed to represent 
nuclear black holes.
Most of the boxlike orbits are found to be stochastic,
with mean Liapunov times that are $3-6$ 
times the period of the long-axis orbit.
In models with large cores or small black holes, the 
stochastic orbits mimic regular box orbits for hundreds 
of oscillations at least.
However a small core radius or significant black hole mass causes 
most of the stochastic orbits to diffuse through phase space on 
the same timescale, visiting a significant fraction of the volume 
beneath the equipotential surface.
Some stochastic orbits, with initial conditions lying close to 
those of regular orbits, remain trapped in all models.

We estimate timescales for chaotic mixing in the more strongly 
stochastic models by evolving ensembles 
of $10^4$ points until their distribution reaches a nearly steady 
state. 
Mixing initially takes place rapidly, with characteristic times 
of $10-30$ dynamical times, as the phase points 
fill a region similar in shape to that of a box orbit.
Subsequent mixing is slower, with characteristic times of hundreds 
of orbital periods.
Mixing rates were found to be enhanced by the addition of 
modest force perturbations, and we propose that the stochastic parts of 
phase space might be efficiently mixed during the early phases of 
galaxy formation when such perturbations are large.
The consequences for the structure and evolution of elliptical galaxies are 
discussed.

\end{abstract}

\section {Introduction}

Elliptical galaxies appear to be smooth and well-ordered systems, and 
this apparent regularity is often taken as evidence for an 
underlying mathematical simplicity.
Jeans (1915) first showed how to construct analytic models of
time-independent galaxies based on forms for the gravitational potential
that support only regular, i.e. non-chaotic, stellar orbits.
Jeans's theorem has motivated a large number of studies based on 
assumed forms for the gravitational potential in which the motion 
is characterized by two or more global invariants.
For instance, all the trajectories in a spherically-symmetric 
potential respect four isolating integrals of the motion, and in 
axisymmetric potentials both the energy $E$ and angular momentum 
about the symmetry axis $L_z$ are globally conserved.
Motion in a spherical potential is therefore fully regular, and
any chaotic orbits in an axisymmetric potential must remain
confined to a phase space region of constant $E$ and $L_z$.
It is common practice in modelling axisymmetric systems to assume  
that the phase space density is constant on surfaces of constant $E$ and 
$L_z$ so that the chaos, if present, is simply ignored.

This picture changes somewhat when we consider less symmetrical 
forms for the potential.
Motion in the potential of an ellipsodially stratified 
mass model can be fully regular, as shown by Kuzmin 
(1973) and de Zeeuw \& Lynden-Bell (1985), 
but Kuzmin's model does not mimic very well 
the distribution of mass or light in real galaxies.
In more general triaxial models, some orbits appear to retain 
three integrals of the motion while others become 
irregular, conserving only the energy
(Merritt 1980; Udry \& Pfenniger 1988).
In a pioneering study, Goodman \& Schwarzschild (1981) showed 
that a significant fraction of the orbits in the triaxial potential 
generated by Hubble's (1930) density law are stochastic.
However they found that the chaos was of little 
consequence for the orbital motion, since stochastic orbits appeared to 
behave very much like regular orbits over timescales of 
$10^2$ orbital oscillations.
They coined the term ``semi-stochasticity'' to describe this 
phenomenon.
Since Goodman \& Schwarzschild's study, most reviews of 
elliptical galaxy dynamics have 
emphasized the near-regularity of the stochastic motion in 
triaxial potentials (e.g. Schwarzschild 1987; Gerhard 1993; 
de Zeeuw 1994).

Hints that chaos might play a larger role in elliptical 
galaxy dynamics appeared in Schwarzschild's (1993) study of 
scale-free, $\rho\propto r^{-2}$, triaxial models.
Schwarzschild found that most of the boxlike orbits in
each of six scale-free models were stochastic.
The exceptions were orbits lying near to stable periodic orbits,
or ``boxlets,'' that avoided the center.
At the same time, it became clear that the distribution 
of light in early-type galaxies is better described by
Schwarzschild's scale-free models than by the models of Hubble 
(1930) or Kuzmin (1973) with their large, constant-density cores.
High-resolution observations 
revealed that elliptical galaxies never have cores; 
instead, the luminosity density always increases monotonically toward the 
center (Crane et al. 1993; Ferrarese et al. 1994; 
Moller, Stiavelli \& Zeilinger 1995; Lauer et al. 1995).
Even elliptical galaxies that were once thought to have 
well-resolved cores, like M87, are now known to contain 
cusps with a weak, power-law dependence of density on radius
(Merritt \& Fridman 1995; Gebhardt et al. 1996).
In addition, there is increasingly strong evidence for central 
mass concentrations, possibly supermassive black holes, at the centers 
of many early-type galaxies (Ford et al. 1994; Miyoshi et al. 1995).
Some nearby galaxies like M32 exhibit both steep central cusps 
and pointlike central mass concentrations (Tonry 1987).

The motion of boxlike orbits in triaxial models with central mass 
concentrations is often strongly chaotic.
Gerhard \& Binney (1985) investigated the two-dimensional motion of 
stars in potentials with central point masses and steep density 
cusps.
They found that central singularities can subject stars on 
boxlike orbits to deflections that destroy their nonclassical 
integrals of motion.
The evolution of such an orbit can be described as a series of 
near-random transitions from one box orbit to another.
Merritt \& Fridman (1996) computed libraries of orbits in two 
triaxial models with Dehnen's (1993) density law:
\begin{equation}
\rho(m)=\rho_0 
m^{-\gamma}(1+m)^{-(4-\gamma)}, \ \ \ m^2={x^2\over a^2}+{y^2\over 
b^2}+{z^2\over c^2}
\end{equation}
with $\gamma = 1$ (``weak cusp'') and $\gamma=2$ (``strong 
cusp'').
They found a similar, large fraction of stochastic orbits in both 
potentials but the behavior of these orbits was different 
in the two cases.
In the weak-cusp potential, almost all of the stochastic orbits 
mimicked regular boxlike orbits for more than $10^2$ oscillations -- 
roughly the same behavior seen by Goodman \& Schwarzschild (1981) 
in their triaxial model with a large core.
In the strong-cusp potential, however, the stochastic orbits 
appeared to diffuse over the energy 
surface on timescales of only $10^2 - 10^3$ oscillations.
After that time, most of the stochastic orbits in the 
strong-cusp model had reached a time-averaged steady state that was 
approximately the same for all stochastic orbits at a given 
energy -- the orbits had essentially filled the ``Arnold web''
(Arnold 1964).
The replacement of distinguishable stochastic orbits by these time-invariant 
ensembles reduces substantially the freedom to construct 
self-consistent equilbria, and in fact Merritt \& Fridman (1996) could 
find no fully stationary solution corresponding to the strong-cusp model.

These studies suggest that chaos is a common consequence of 
triaxiality,
and furthermore that the timescales over which the chaos 
manifests itself in the orbital motion can be physically interesting 
-- long compared to a crossing time, but often (and especially near the 
center) short compared to a Hubble time.
In this paper we attempt to more accurately calculate these 
timescales.
Our goal is to understand the role that stochasticity 
plays in the structure and evolution of early-type 
galaxies.
If the timescale for chaotic orbits to fill their allowed phase space is
short compared to the age of a galaxy, stochasticity could strongly 
reduce the variety of orbits available for the construction of 
equilibrium galaxies.
This in turn might imply that triaxiality is a rare phenomenon 
(Schwarzschild 1981).
On the other hand, if chaotic timescales are similar to or 
greater than a galaxy 
lifetime, triaxial galaxies might be slowly evolving as the 
stochastic orbits continue to diffuse --- an equally interesting 
possibility.

We investigate these questions by integrating ensembles of orbits 
in a family of triaxial models that mimic 
the distribution of light in early-type galaxies.
Our family (\S2) contains Kuzmin's integrable model as a special case.
By varying a free parameter, $m_0$, we can change the size of the 
constant-density region near the center; for $m_0=0$, our models 
have a $\rho\propto r^{-2}$ central cusp similar to those observed 
in some early-type galaxies.
We are thus able to study the way in which departures from 
perfect integrability ($m_0 = 1$) induce stochasticity in the motion.
We find (\S3) that the number of stochastic orbits is not a 
strong function of $m_0$, and that the typical Liapunov exponent 
is $0.1-0.3$ in units of the inverse dynamical time (defined 
as the full period of the long-axis orbit).
Some of the stochastic orbits appear to mimic regular orbits for 
hundreds of oscillations or even longer.
This ``trapping'' affects virtually all of the stochastic orbits 
in the potentials with large $m_0$, but is much less important as 
$m_0\rightarrow 0$, and most of the stochastic orbits in the 
cuspy models behave ergodically over astronomical timescales 
(\S4).

Although our calculations are restricted to motion in a single family of 
triaxial models, our results are relevant to a broader question.
How do galaxies reach a steady state?
Jeans's theorem requires that the phase-space density $f$ 
be constant on the tori that define the range of motion of the 
regular orbits.
But it is well known that the evolution of a collisionless 
ensemble of stars is never toward the uniform population of 
phase space demanded by Jeans's theorem.
Any initially compact group of phase points gets drawn out into a 
filament of ever-decreasing width as they move 
independently in response to the gravitational force.
Observed with infinite resolution, the phase space 
occupied by these points becomes increasingly striated, not more uniform.
The most we can hope for is that the coarse-grained phase space density
will approach a constant value within some region; 
but even this outcome is not guaranteed by any general property of 
Hamilton's equations.

One mechanism by which an ensemble of regular 
trajectories can evolve to a coarse-grained steady state is phase mixing.
A simple example of phase mixing is a set of stars on circular 
orbits with the same initial phase, $f_0(r,\phi)=\delta(\phi-\phi_0)$,
$r_1<r<r_2$.
If the circular frequency $\Omega$ is a function of radius,
this sliver will be wound into a filament (Figure 1a); the Fourier 
coefficients of the distribution in configuration space will be
\begin{eqnarray}
	A_k 	&\propto& \int_{r_1}^{r_2} dr\ r^2\int_0^{2\pi} d\phi\ 
	\cos{k\phi}\ \delta\left[\phi-\Omega(r)t\right] \\
		&\propto& \int_{r_1}^{r_2} dr\ r^2\cos\left[k\Omega(r) 
		t\right].
\end{eqnarray}
For any smoothly-varying $\Omega(r)$, the Riemann-Lebesgue theorem 
guarantees that at late times $A_k\rightarrow 0$ for $k\ne 0$.
Thus the coarse-grained density, i.e. the density averaged over 
some radial interval, tends to a value that is independent of 
angle even though the density of points on any particular orbit 
is nonzero only at a single value of $\phi$.

Phase mixing undoubtedly plays a role in the approach of 
collisionless stellar systems to equilibrium.
But phase mixing is an intrinsically slow process -- in fact it 
has no well-defined timescale.
The rate at which a group of phase points shears depends on the range 
of orbital frequencies in the group.
If the maximum and minimum frequencies are $\Omega_1$ and 
$\Omega_2$ respectively, we expect phase mixing to take place on a 
timescale of order $(\Omega_1-\Omega_2)^{-1}$.
This timescale is never less than a dynamical time and can be 
much longer.
For instance, near the half-mass radius $r_e$ of a de Vaucouleurs-law 
galaxy, the phase mixing time $2\pi/(\Omega_1-\Omega_2)$
for two stars on circular orbits with separation $\Delta r$
is roughly $0.85/(\Delta r/r_e)$ times the orbital period.
Inhomogeneities on a scale of $0.1 r_e$ -- roughly 100 pc in a 
real galaxy -- damp out over $\sim 10$ dynamical 
times, and shorter length scales reach equilibrium even more 
slowly.
In the limit $\Omega_1\rightarrow\Omega_2$ there is no phase mixing.
This is the case for a set of points that is restricted to a single invariant 
torus: the ensemble simply translates, with fixed shape, around the 
torus as the phase points move with the same fixed set of frequencies.

Phase mixing is also limited in the sense that it can only 
reshuffle phase points within a narrow region, since 
by definition the trajectories are confined to invariant tori.
Mixing in a system with a time-evolving 
potential, or in a system containing irregular orbits, 
must operate via some different mechanism.

Returning to the example discussed above, we can ask: what
property of the motion guarantees the vanishing, after long 
times, of the Fourier components of $f$?
The periodicity of the motion with respect to time is clearly 
sufficient for this purpose, but it is not necessary.
One could imagine more general flows for which the coarse-grained 
density converges to a form that is independent of angle.
In fact any motion which has the property that the average time 
spent in any phase interval $d\phi$ is proportional to $d\phi$ will do 
the trick, since at late times $\phi$ will have no preferred value
at any radius.
This includes flows that are discontinuous, or even random, with
respect to time.
We will follow the now-standard practice of calling ``ergodic''
any motion that spends equal amounts of time, 
on average, in equal phase-space volumes (defined with respect to
the invariant measure).\footnote[2]{The term ergodic was originally
used to describe motion that visits every point on the energy surface.
Modern usage recognizes the fact that this stringent condition is
rarely satisfied in real systems, and hence that ergodicity is more 
usefully defined over some limited part of the energy surface, such 
as an invariant torus (Lichtenberg \& Lieberman 1983, p.260.}
Ergodicity is a non-trivial property of dynamical systems; it is 
characteristic of quasi-periodic motion, i.e. regular orbits
(Arnold 1989, p. 287), but other types of motion can be ergodic as well.

While ergodicity is a necessary property of the motion if 
phase mixing is to produce a steady state, ergodicity by itself is a 
condition only on the time-averaged behavior of a trajectory.
Relaxation implies more in general: 
not only should the entire past of a given phase point cover the phase
space uniformly, but so also should the present of any neighborhood of the 
original point.
In other words, any small patch of phase space should evolve
in such a way that it uniformly covers, at a {\it single} 
later time, a much larger region.
Motion that is restricted to a single torus, 
while ergodic, does not have this stronger property, since a 
collection of phase points simple translates unchanged
around the torus.
Phase mixing, which involves points on different tori, does 
produce a weak sort of relaxation but only within the narrow 
region allowed by the integrals of motion.

Dynamicists define as ``mixing'' any system which exhibits this
stronger form of relaxation.
A standard definition of a mixing system (e.g. Lichtenberg \& 
Lieberman 1983, p. 268) is one in which any portion of the phase space, 
however small, tends to be uniformly distributed over the energy 
surface (or some subspace defined by the integrals of motion) 
as the time increases indefinitely (Fig. 1b).
Mixing systems are always ergodic (Arnold \& Avez 1968), 
but the converse is not true; for instance, motion on the torus 
is ergodic but not mixing.
Mixing systems have the further property that they relax:
the density in a mixing system evolves 
toward a constant, coarse-grained value at all accessible phase 
space points.
The trajectories in mixing systems are often stochastic,
since stochasticity guarantees the drastic loss of 
correlations that is required if relaxation is to erase memory of
the initial state.
In fact the canonical examples of mixing systems, such as the 
hard-sphere gas (Sinai 1976), are fully chaotic systems.
Mixing also has an associated timescale (Krylov 1979); in this sense 
too it differs from phase mixing for which no 
characteristic time can be defined.

A mixing flow has many of the properties that we 
associate with the approach to equilibrium of collisionless 
stellar systems.
The mixing property has been rigorously proved only for a small set of 
idealized, strongly chaotic dynamical systems 
(Ott 1993, p. 257).
In potentials containing both regular and stochastic orbits, like 
those of galaxies, we intuitively expect that the stochastic
orbits -- which are effectively random in their long-time behavior 
-- will be both ergodic and mixing over the bounded portion of the
phase space for which they exist.
However the time required for the mixing to produce an equilibrium
state might be long, especially if the stochastic motion is hampered
by the presence of invariant tori and ``cantori.''
Although a proof of mixing in realistic galactic potentials 
would be extremely difficult, we can still hope to observe the signature of 
mixing behavior -- approach to a characteristic, coarse-grained 
equilibrium on some well-defined timescale -- in numerical 
experiments.
This we do, in \S5.

In a real galaxy, the approach to a steady state 
takes place against the backdrop of an evolving potential.
Mixing should be more efficient in this case,
since energy is no longer an invariant of the motion and stars 
are free to move throughout the entire phase space.
Furthermore a potential that is changing in a complicated way 
with time might induce chaotic motion, or something 
like it, in the particle trajectories.
Although we do not attempt here to simulate this more complicated 
process, we do show in \S5 that the addition of 
random force perturbations can enhance the mixing rates of 
stochastic ensembles.
This leads us to propose that rapidly-varying forces
during galaxy formation might cause the stochastic parts of phase 
space to be strongly mixed by the time the gravitational 
potential settles down.
The replacement of distinguishable stochastic orbits by a smaller 
number of invariant ensembles would reduce the freedom to 
construct self-consistent equilibria; some possible consequences 
for real galaxies are discussed in \S6.

\section{A Family of Triaxial Models}

In this section we identify a family of mass models that resemble real  
elliptical galaxies and bulges and calculate their gravitational potentials.
Our starting point is the Perfect Ellipsoid (Kuzmin 1973; de 
Zeeuw \& Lynden-Bell 1985), with mass distribution
\begin{equation}
\rho(m) = {\rho_0\over(1+m^2)^2}. 
\label{perfect}
\end{equation}
Motion in the gravitational potential of the Perfect Ellipsoid
is characterized by three global integrals, i.e. the motion is 
completely regular.
All orbits fall into one of four families: the boxes, and 
three families of tubes (Kuzmin 1973; de Zeeuw 1985).
Tube orbits circulate either around the long ($x$) or short ($z$) 
axis of the figure, and conserve a quantity analogous to the 
angular momentum about that axis.
They therefore avoid the center.
Box orbits have filled centers and touch the equipotential 
surface at eight points, one in each octant.

Kuzmin's density law was arrived at via mathematical 
manipulations and it is not surprising that the Perfect Ellipsoid 
bears little resemblance to the distribution of light or mass in 
real elliptical galaxies.
The discrepancy is particularly great near the center, where 
Kuzmin's law predicts a large, constant-density core.
The luminosity densities in real elliptical galaxies and bulges 
are always observed to rise monotonically at small radii 
(Ferrarese et al. 1994; Moller, Stiavelli \& Zeilinger 1995; 
Lauer et al. 1995).
The steepest profiles are seen in low-luminosity galaxies like M32, 
while brighter galaxies like M87 have surface brightness profiles 
that look superficially core-like (Kormendy et al. 1995).
However Merritt \& Fridman (1995) showed that even these ``core'' 
galaxies have power-law cusps in the luminosity density.
These galaxies appear to have cores only because the logarithmic 
slopes of their density cusps are less than $-1$, and a shallow 
power law cusp produces a gently curving surface brightness 
profile when seen projected through the outer layers of the galaxy
(Dehnen 1993, Fig. 1).

We would like to find a simple mathematical expression for the 
mass distribution that reproduces, at least qualitatively, the 
distribution of light seen in real galaxies.
Ideally our expression should contain Kuzmin's law as a special case 
so that we can study the way in which departures 
from the Perfect law generate irregular motion.
In addition, the gravitational forces generated by our model should 
be simple to calculate in order that the orbit integrations not 
be too slow.

A mass model that satisfies each of these requirements is
\begin{equation}
\rho(m) = {\rho_0m_0^2\over (1+m^2)(m_0^2+m^2)} , \quad 0 \le
m_0 \le 1.
\label{density}
\end{equation}
The central density $\rho_0$ is related to the total mass
$M$ by the expression
$$\rho_0 = {{M(1+m_0)}\over{2\pi^2 abcm_0^2}}.$$
The density profile breaks into three regions (Fig. 2).
At large radii, $m\gg 1$, the density falls off as $m^{-4}$, as 
in the Perfect law.
At intermediate distances from the center, $m_0\lap m\lap 1$,
$\rho\sim m^{-2}$.
Near the center, $m\lap m_0$, the density becomes constant.
For $m_0 =1$ the profile reduces to Kuzmin's law, while 
for $m_0=0$ it has a $\rho\propto m^{-2}$ 
central density cusp like those observed in some
elliptical galaxies.
Thus varying $m_0$ from 1 to 0 takes one from a fully integrable 
but nonphysical model, to a realistic but (as we will see) 
strongly nonintegrable model.

We do not have the freedom with our family of models to adjust the slope of 
the central density cusp to any power-law index.
We are therefore not able to reproduce the shallow power-law 
cusps observed in luminous elliptical galaxies.
However in a galaxy where the density increases near the center as 
a power law of index $\gamma$, the radial force reaches a maximum 
value at some nonzero radius for $\gamma<1$, and we might expect the 
motion in such a galaxy to be crudely reproducable with our models if 
$m_0$ is chosen to have an appropriate value.
For instance, in a galaxy with Dehnen's (1993) density law, 
$\rho\propto r^{-\gamma}(a+r)^{\gamma-4}$, 
the radial force peaks at $r=a(1-\gamma)/2$ for $\gamma<1$.
Thus from a dynamical point of view such a galaxy has a ``core'' 
even though the density is diverging near the center.

The gravitational potential generated by an 
ellipsoidally-stratified mass distribution is 
\begin{equation}
{\bf \Phi(x)} = - \pi G abc \int_0^{\infty}{{[\psi(\infty) -
\psi(m)] du}\over{\sqrt{(u+a^2)(u+b^2)(u+c^2)}}},
\label{genpotential}
\end{equation}
(Chandrasekhar 1969, Theorem 12), where
$$
\psi(m) = \int_0^{m^2}{\rho(m^{\prime 2})dm^{\prime 2}}
$$
and
$$
 m^2(u) = {{x^2}\over{a^2+u}} + {{y^2}\over{b^2+u}} +
{{z^2}\over{c^2+u}}.
$$
For the density distribution (5), we find
\begin{equation}
\psi(m) = {1\over{2\pi abc(1-m_0)}} \left[\log(1+{m^2/m_0^2}) -
\log(1+m^2)\right]
\end{equation}
and the potential becomes
\begin{equation}
{\bf\Phi(x)} = {\bf \Phi_0} - {1\over{2\pi(1-m_0)}}
\int_0^{\infty}{\left[\log(1+m^2) - \log(1+m^2/m_0^2)\right]\over
\sqrt{(u+a^2)(u+b^2)(u+c^2)}} du
\label{potential}
\end{equation}
where
\begin{equation}
{\bf \Phi_0} = {2\log m_0\over \pi(1-m_0)} R_F(a^2,b^2,c^2)
\label{potconst}
\end{equation}
and $R_F$ is Carlson's incomplete elliptic integral of the first 
kind.
We have adopted units in which the total
mass $M$, the gravitational constant $G$, and the long axis
length $a$ are equal to one.

Gravitational forces can be computed by taking the gradient 
of Eq. (8).
The expressions so obtained are expensive to compute numerically, 
since they contain an integral which must be evaluated anew at every point 
along the trajectory.
A better way to proceed is to make use of the fact that the 
density (5) can be written 
\begin{equation}
 \rho(m) = {1\over{2\pi^2 abc(1-m_0)}}\left[{1\over{m^2+m_0^2}}
- {1\over{m^2+1}}\right].
\label{std_density}
\end{equation}
De Zeeuw and Pfenniger (1988) show that the gravitational potential 
generated by a mass density of the 
form $\rho=\rho_0/(m_0^2 + m^2)$ can be expressed in 
terms of confocal ellipsoidal coordinates $(\lambda, \mu, \nu)$, 
defined as the three roots for $u$ in the equation $m^2(-u) = -1$. 
%These roots satisfy the conditions $c^2 < \nu \le b^2 < \mu \le a^2 \le
%\lambda$ (cf. de Zeeuw 1985).
On transforming to ellipsoidal coordinates we have
\begin{equation}
 1 + m^2(u) = {{(\lambda + u)(\mu + u)(\nu + u)}\over
{(a^2 + u)(b^2 + u)(c^2+u)}},
\end{equation}
and the potential becomes
\begin{equation}
{\bf\Phi(x) = \Phi_0}- {1\over{2 \pi (1-m_0)}} \left[G(\lambda) + G(\mu) +
G(\nu) - m_0\left( H(\lambda^{\prime}) + H(\nu^{\prime}) +
H(\mu^{\prime})\right)\right].
\end{equation}
Here $(\lambda, \mu, \nu)$ and $(\lambda^{\prime}, \mu^{\prime},
\nu^{\prime})$
are separate sets of ellipoidal coordinates corresponding 
respectively to the 
axis lengths $(a, b, c)$ and ($ am_0, bm_0, cm_0$),
and
\begin{eqnarray}
 G(\tau) & = & \int_0^{\infty} {{\log(\tau+u) du}\over
{\sqrt{(a^2+u)(b^2+u)(c^2+u)}}}, \\
 H(\tau^{\prime}) & = & \int_0^{\infty} {{\log(\tau^{\prime}+u) du}
\over{{\sqrt{(a^2m_0^2 + u)(b^2m_0^2 + u)(c^2m_0^2 + u)}}}}.
\end{eqnarray}
Furthermore it can be shown that
\begin{equation}
 m_0 H(\tau^{\prime}) = G(\tau^{\prime}/m_0^2) + {\rm constant}
\end{equation}
so that
\begin{equation}
{\bf \Phi(x)} = {\bf\Phi_0} -
{1\over{2\pi(1-m_0)}}\left[G(\lambda) + G(\mu) + G(\nu) -
G(\lambda^{\prime}/m_0^2) - G(\mu^{\prime}/m_0^2) -
G(\nu^{\prime}/m_0^2)\right].
\end{equation}

This expression for the gravitational potential is not necessarily 
faster to evaluate numerically than the one based on 
Chandrasekhar's formula in Cartesian coordinates.
However, the gravitational forces may now be expressed as sums 
of terms like
\begin{equation}
{(\lambda-b^2)(\lambda-c^2)\over (\lambda-\mu)(\lambda-\nu)} 
R_J(a^2,b^2,c^2,\lambda),
\end{equation}
with $R_J$ Carlson's third incomplete elliptic integral.
The functions $R_J$ depend on a single argument and may be 
approximated via splines at the start of the integration.
In this way, the calculation of orbits can be speeded up by 
roughly an order of magnitude without a significant loss in 
accuracy.
The details are presented in the Appendix.

There is increasingly strong evidence for central singularities, 
possibly supermassive black holes, at the centers of a few
early-type galaxies (Ford et al. 1994; Miyoshi et al. 1995).
It is possible that most or all elliptical galaxies contain such 
black holes with masses in the range $0.1-1\%$ the total stellar mass
(Kormendy \& Richstone 1995).
Accordingly, we also investigate the orbital motion in triaxial 
models containing central point masses.
The masses of these ``black holes,'' $M_{BH}$, will henceforth be expressed
as fractions of the galaxy mass.

\section {Liapunov Exponents}

In this section we compute the simplest index 
of stochasticity, 
the Liapunov exponents, for a large number of orbits in the 
gravitational potential just defined.
The Liapunov exponents are interesting in their own right, as 
measures of the rate of divergence of nearby trajectories and as 
diagnostics for separating regular from stochastic orbits.
But we also expect the Liapunov exponents to be related to 
the rate of mixing, since mixing is driven by the 
spreading of trajectories.

The phase space of the potential (8) 
is partly regular and partly stochastic.
The majority of stochastic trajectories in a triaxial potential
have the property that 
they touch the surface $\Phi({\bf x})=E$ at a set of stationary 
points (Schwarzschild 1993; Merritt \& Fridman 1996).
Our initial conditions were therefore taken to be on one octant 
of the equipotential surface, with zero velocity.
We selected the starting points of each isoenergetic ensemble
of orbits on a regular grid of 192 points distributed over the 
equipotential surface, as described by Merritt \& Fridman (1996).

The Liapunov exponents of a trajectory are the mean exponential
rates of divergence of trajectories surrounding it.
In a three-degrees-of-freedom system there are six Liapunov 
exponents for every trajectory, corresponding to the six 
dimensions of phase space.
The exponents come in pairs of opposite sign;
of the three independent exponents, one -- corresponding to 
displacements in the direction of the motion -- is always zero.
We are thus left with two independent exponents, $\sigma_1$ and 
$\sigma_2$.
These may be seen as defining the time-averaged 
divergence rates in two directions orthogonal to the trajectory.
For a regular orbit, $\sigma_1=\sigma_2=0$; for a stochastic 
orbit, at least one (and typically two) of these exponents is nonzero.

Liapunov exponents are defined as limiting values 
over an infinite time interval (e.g. Lichtenberg \& Lieberman 
1983, p. 264), 
and hence are impossible to calculate via any finite numerical scheme.
We computed approximations to the Liapunov exponents by 
integrating orbits for long periods, up to $10^4$ dynamical 
times $T_D(E)$, defined as the full period of the $x$-axis orbit 
of energy $E$.
We used the Gram-Schmidt orthogonalization
technique described by Benettin et al. (1980) in an 
implementation developed by the Geneva Observatory group
and kindly made available by Dr. St\'ephane Udry.
The evolution of the perturbed orbits is determined by the second
derivatives of the potential with respect to position; these
expressions are given in the Appendix.
The time required to integrate one orbit and its six
perturbation orbits for $10^4$
orbital times and compute the Liapunov exponents was 
about 80 minutes on a DEC Alpha 3000/700 workstation.
Henceforth we will use the term ``Liapunov exponents'' to refer to 
these finite-time, numerical approximations to the true 
exponents.

We fixed the axis ratios of our triaxial model to $c/a=T=0.5$, 
where $T=(a^2-b^2)/(a^2-c^2)$ is the ``triaxiality index''; 
our choice for $T$ corresponds to ``maximum triaxiality.''
We chose three values for the core radius, $m_0=\{0.1, 0.01, 0.001\}$,
and three values for the mass of the central ``black hole,'' 
$M_{BH}=\{1, 3, 10\}\times 10^{-3}$.
For comparison, the ratio of black hole mass to luminous galaxy 
mass is thought 
to be about $5\times 10^{-3}$ for M87 and $2.5\times 10^{-3}$ for 
M32 (Kormendy \& Richstone 1995).
All orbits had an energy equal to that of the long-axis orbit 
that just touches the ellipsoidal shell dividing the model into 
two equal-mass parts.
The amplitude of this orbit is roughly 1 in all of the models 
considered here.

Figures 3-5 shows histograms of $\sigma_1$ and $\sigma_2$ for 
various combinations of $m_0$ and $M_{BH}$.
Each plot contains three curves corresponding to integration times of 
$10^2$, $10^3$ and $10^4 T_D$ for the same set of 192 orbits.
As the integration time increases, the separation of the orbits 
into two groups becomes apparent.
The Liapunov exponents of the regular orbits, the ``boxlets'' 
(Miralda-Escude \& Schwarzschild 1989), 
lie in a narrow peak near zero.
The Liapunov exponents of the stochastic orbits show a larger 
spread, but the spread decreases with time and the mean value 
does not change very much after $10^3T_D$.

We expect that every stochastic orbit at a given 
energy moves in the same stochastic ``sea,'' interconnected via the 
Arnold web (Arnold 1964).
One can begin to identify, after $10^4$ dynamical times, the 
unique numbers $\sigma_1(E)$ and $\sigma_2(E)$ that characterize 
the stochastic motion in this sea.
Table 1 gives Liapunov exponents in units of $T_D^{-1}$ 
at $t=10^4 T_D$ for various combinations of $m_0$ and $M_{BH}$.
These values are simple averages over the 
subset of orbits that lie within the stochastic part of the 
histogram, defined to be the part to the right of the narrow
peak defining the regular orbits.
Also given there are the dispersions of the $\sigma$ values about 
their means, and the number of orbits, out of the total 192, that are 
stochastic.

A large fraction of the non-tube orbits are stochastic, and this fraction 
is not tremendously dependent on $m_0$ or $M_{BH}$.
Even for $m_0=0.1$ and $M_{BH}=0$ --- not greatly different from the 
Perfect Ellipsoid --- about half of the equipotential surface 
generates stochastic orbits, and this fraction increases to $\sim 
150/192$ when $m_0$ has the physically more realistic values of 
$10^{-2}$ or $10^{-3}$.
The addition of a central point mass does not increase the 
fraction of stochastic orbits very much.
The Liapunov exponents for the most stochastic model considered 
here, $m_0=10^{-3}$ and $M_{BH}=0.01$, are 
$\sigma_1\approx 0.32$ and $\sigma_2\approx 0.13$ in units of the 
inverse dynamical time.
The most nearly regular model, $m_0=0.1$ and $M_{BH}=0$, has 
mean Liapunov exponents that are only a factor of $\sim 2-3$ lower than
these extreme values.

Roughly speaking, then, divergence between nearby stochastic orbits 
takes place on timescales that average 3-6 times the 
dynamical time in all of our models.

We found that $\sigma_2$ correlated strongly with $\sigma_1$ in our 
ensembles; there were few if any orbits with large $\sigma_1$ and 
small $\sigma_2$.
A vanishing $\sigma_2$ would imply the existence of an isolating integral 
in addition to the energy.
It appears that most or all of the stochastic orbits in our model potentials
respect only the energy integral.

\section {Trapped Orbits}

Figures 3-5 give hints of a second, longer timescale associated with the 
stochasticity.
After an infinite time (according to Arnold's conjecture) 
all of the stochastic orbits in an 
isoenergetic ensemble should have exactly the same values of $\sigma_1$ and 
$\sigma_2$.
While our results are consistent with this conjecture, 
Figures 3-5 show that the spread in the Liapunov exponents 
at a given energy decreases very slowly with time, 
remaining appreciable even after $10^3 T_D$.
Much of the spread is due to extended tails in the histograms 
toward low values of $\sigma_1$ and $\sigma_2$.
The slow rate at which the Liapunov exponents evolve toward a
common value suggests that many stochastic orbits
fail to explore the full stochastic phase space in an effectively 
ergodic manner.
Apparently, this trapping can persist for periods of 
time much longer than $\sigma_1^{-1}$.
Inspection of the configuration-space plots confirms this 
expectation.
Many of the stochastic orbits are found to mimic regular boxlike 
orbits for hundreds of oscillations before suddenly changing to a different 
boxlike shape, etc.
Similar behavior has been noted by many authors (e.g. 
Goodman \& Schwarzschild 1981; Binney 1982a).

The confinement of stochastic orbits to limited parts of phase 
space over long periods of time is due to the fact that our phase  
space is ``decomposable,'' i.e. contains both regular and stochastic 
parts.
In a decomposable phase space, some of the stochastic trajectories 
travel close to the imbedded tori and can become stuck there for a long 
time (e.g. Karney 1983).

We investigated the degree of trapping via the scheme described 
by Goodman \& Schwarzschild (1981).
The surface of section $E=E_0$, $x=y=z=0$ is intersected by the 
majority of boxlike orbits, excluding only those that lie close 
to a stable resonance that avoids the center.
An orbit respecting two integrals of the motion in addition to 
the energy will pass through this point with at most a finite 
set of velocity vectors, while an orbit respecting only the 
energy integral will not be constrained in the direction of its 
velocity.
A trapped stochastic orbit may be distinguished from a more 
freely-moving one by inspecting the fraction of the 
velocity-space sphere that it covers after successive passages 
through the central point.

We constructed surfaces of section for each of the orbits in 
our ensembles by integrating them for a time interval of 500 $T_D$
and recording their velocities at the central crossings.
In practice, this meant recording passages through the sphere 
$(x^2+y^2+z^2)^{1/2}<0.03$.
These velocities were then mapped onto one octant of the 
velocity-space sphere, as in Goodman \& Schwarzschild (1981).

In the potential with the largest core and no black hole, 
$m_0=0.1$ and $M_{BH}=0$, 
the stochastic orbits all remained confined to small regions in 
this surface of section.
The largest scatter was produced by stochastic orbits with starting points 
near the $y-z$ plane and the $z$ axis, but even these orbits covered 
an area only $\sim 4-6$ times that of a typical regular orbit on 
the surface of section.
This behavior is similar to that described by Goodman \& 
Schwarzschild (1981) for motion in a Hubble-law 
potential.
It is also consistent with the histograms of Figure 3, 
which show considerable spread in the Liapunov 
exponents of the orbits in this model even after $10^3 T_D$.

When $m_0$ was decreased to $10^{-2}$ or $10^{-3}$, however, the 
scatter in the surface of section increased dramatically.
Figure 6a shows the velocities at central crossings for three 
orbits with similar starting points in the three potentials 
defined by $m_0=10^{-1}, 10^{-2}$ and $10^{-3}$, with $M_{BH}=0$.
In the potential with the smallest core, the orbit fills a 
large fraction of the surface of section during the period of 
integration; the only indication that the behavior is not fully
random is a clustering of the points in two or three 
sub-regions of the octant.
This nearly-random behavior on the surface of section 
was typical of most of the stochastic 
orbits in the potential with $m_0=10^{-3}$.
A similar, though smaller, degree of scatter was produced by 
stochastic orbits in the $m_0=10^{-2}$ potential.

These results suggest that 
the degree of trapping of stochastic orbits -- unlike the total 
{\it number} of stochastic orbits -- depends strongly on 
the departure of the potential from integrability.
Triaxial potentials with large cores are effectively integrable, not 
because the number of stochastic orbits is small, but because 
these orbits behave like regular orbits over astronomically 
interesting timescales.
The high degree of trapping observed by Goodman \& Schwarzschild (1981)
was apparently due to their choice of a triaxial potential with a large
core.

Increasing the black hole mass had a similar effect to decreasing 
$m_0$.
Figure 6b shows surfaces of section for three orbits in the 
potential with $m_0=0.1$, and $M_{BH}=0.001, 0.003$ and $0.01$.
Most of the stochastic orbits in the model with the smallest black hole act 
only slightly less trapped than in the corresponding model with $M_{BH}=0$.
However a larger $M_{BH}$ produces more scatter, and when
$M_{BH}=0.01$, many of the stochastic orbits appear to scatter 
nearly randomly over the surface of section.
Thus, either a small core ($m_0\lap 10^{-2}$) or a massive central 
singularity ($M_{BH}\gap 0.003$) causes the majority of the 
stochastic orbits to behave ergodically over astronomical 
timescales.

But even in the potentials with small $m_0$ and large $M_{BH}$, a 
certain fraction of the stochastic orbits were found to remain constrained 
to small areas on the surface of section.
Among orbits from the same isoenergetic ensemble, we found that 
the amount of scatter on the 
surface of section correlated fairly well with $\sigma_1$ and 
$\sigma_2$.
Trapped stochastic orbits had Liapunov exponents that were 
smaller than average, usually lying between the two peaks in the 
histogram associated with 
the regular orbits and the stochastic orbits .
We found that the separation was particularly clear in 
histograms of the ``Kolmogorov entropy'' $h_K = \sum_{i=1}^3 \sigma_i$, 
the sum of the three Liapunov exponents.
(Since $\sigma_3\approx 0$, $h_K$ is essentially the sum of the 
two largest exponents.)
For instance, when Liapunov exponents were computed over a time 
interval of $10^3 T_D$, most of the trapped stochastic orbits 
were found to have values of $h_K$ that were less than $\sim 0.5$ 
times the maximum value for the ensemble.
Figure 7 is a map of the starting points of trapped stochastic 
orbits defined in this way for two models: $m_0=10^{-3}, 
M_{BH}=0$ and $m_0=10^{-1}, M_{BH}=3\times 10^{-3}$.
The trapped orbits are not distributed randomly on these plots, 
but instead tend to cluster around the starting points of the 
regular orbits.

Since trapped orbits can mimic regular orbits for astronomically 
interesting timescales, they might be useful building blocks for real 
galaxies.
Schwarzschild (1993), in a study of self-consistent models of galactic 
haloes, allowed each of the stochastic orbits to have its own 
weight -- in effect assuming that every stochastic orbit was trapped.
Merritt \& Fridman (1996), in their ``fully mixed'' models, made 
the opposite assumption, i.e. that all stochastic orbits belong 
to invariant ensembles.
The true situation lies between these extremes: trapped 
stochastic orbits can legitimately be treated like regular orbits, 
at least over appropriate timescales, but the remainder of the 
stochastic orbits at a given energy behave ergodically and 
should be assigned to ensembles with a fixed density distribution.

\section {MIXING}

\subsection{Definitions}

The Liapunov exponents, as well as the scatter on the surface of 
section, are measures of the time-averaged behavior of an orbit.
But nature is less concerned with time averages than with the 
rate of evolution, at a given time, of ensembles.
For instance, a regular orbit covers its invariant torus ergodically in 
a time-averaged sense, but an ensemble of points on the same 
torus does not evolve toward a steady state -- it
simply translates, unchanged, around the torus.
Here we investigate the more interesting sort of relaxation that 
takes place when an ensemble of stochastic trajectories, initially 
confined to a small phase space region, evolves to fill 
(in a coarse-grained sense) the larger region 
accessible to it.
This process is similar to what occurs in a mixing system 
and we will use the term ``mixing'' to describe the 
evolution that we see.
However we emphasize that a rigorous proof that mixing, as  
defined by mathematicians (Arnold \& Avez 1968), 
takes place in our system would be extremely difficult.
Our more modest goal is to show that the signatures of mixing 
behavior -- erasure of correlations and evolution toward 
a coarse-grained steady state -- are present and to estimate the 
timescale over which this evolution occurs.

We need first to define what we mean by a ``fully-mixed'' state,
and then to define a measure of the distance between this
state and the coarse-grained density of the evolving ensemble.
The mixing rate will then be defined as the rate at which this
distance measure approaches zero (e.g. Kandrup \& Mahon 1994).

A stochastic trajectory of energy $E$ moves within a
five-dimensional phase space region.
An obvious definition of the fully-mixed state would be a
constant phase space density throughout this region.
However it is not practical to keep track of the motion within a region
of such high dimensionality, nor can we easily identify the
regular and stochastic parts of phase space.
Instead, we will base our analysis on the three-dimensional, 
configuration space density $\rho({\bf x},t)$ of the evolving ensemble.
The coarse-grained equivalent of $\rho$ will be specified via the occupation
numbers $n_i$ of particles within a set of Cartesian cells.

Since a stochastic trajectory will eventually visit every point in
configuration space that lies within the equipotential surface,
a simple measure of the departure from a fully-mixed state
would be the fraction $F$ of
configuration space cells, lying beneath the equipotential
surface, that contain no particles at a given time.
$F=0$ then corresponds to a fully-mixed state.
This definition ignores the detailed number of particles within the
cells; it indicates only the fraction of accessible volume that is
occupied in a coarse-grained sense.

A better description of a fully-mixed state is the
density corresponding to a uniform population of the energy
surface $E=E_0$, or
\begin{equation}
\rho_{MC}({\bf x})\propto\int\delta(E-E_0)\ d^3v=C\sqrt{E_0-\Phi({\bf
x})},\ \ \ \Phi({x})\le E_0 .
\end{equation}
This ``micro-canonical'' density is still only an approximation
to the fully-mixed state, since no stochastic orbit can sample the
full energy surface -- regions corresponding to the regular
orbits are excluded.
However $\rho_{MC}$ is easy to compute, and in a strongly stochastic system,
we might expect $\rho$ to approach $\rho_{MC}$ fairly closely.

A third, and precise, definition of a fully-mixed state would be the density
reached, after an infinite integration time, by an
ensemble of particles that evolves according to the
equations of motion in stochastic phase space.
We can never compute this density exactly, but we may be able to
approximate it in cases where the mixing is
sufficiently rapid.
%We will call this density $\rho_{FM}({\bf x})$ with the
%understanding that $\rho_{FM}$ can only be computed
%approximately.

The ``distance'' between $\rho({\bf x},t)$ and $\rho_{MC}({\bf
x})$ 
%or $\rho_{FM}({\bf x})$ 
can be defined in various ways.
One way is in terms of the Jaynes entropy (Dejonghe 1987),
\begin{equation}
S(\rho,\rho_0) = -\int \rho\left[\ln\left({\rho\over \rho_0}\right)
-1\right]d{\bf x}
\label{entropy}
\end{equation}
with $\rho$ defined as the probability density function in
configuration space, normalized such that
 $\int \rho({\bf x},t) d{\bf x} =\int \rho_0 d{\bf x}=1$.
In discrete form, the corresponding distance measure $1-S(\rho,\rho_0)$
becomes
\begin{equation}
D_1(n_i,{n_0}_i) = \sum_i n_i\ln\left({n_i\over {n_0}_i}\right),
\label{entdis}
\end{equation}
where $n_i$ is the number density of particles in configuration-space
cell $i$, and ${n_0}_i$ is the number density of particles
predicted by the reference density, or $\rho_{MC}$ 
in the present case. 
The cell densities are normalized such that 
$\sum_i n_i = \sum_i {n_0}_i =1$.
Equation (19) measures the logarithm of the probability
that the
$n_i$ would have been measured if the occupation numbers were generated
from the distribution $n_0$.
$S$ is maximized, $S=1$, when $n_i={n_0}_i$ for all $i$, 
and the fully-mixed state therefore has $D_1=0$.
In the limit $n_i \rightarrow {n_0}_i$, we have $\ln(n_i/{n_0}_i)
 \rightarrow \ln(1+\delta) \rightarrow \delta$
 where $\delta =(n_i-{n_0}_i)/{n_0}_i$.
Thus
\begin{eqnarray}
D_1 \rightarrow \sum_i n_i\left[{n_i\over{n_0}_i} -1\right].
\label{ent_lim}
\end{eqnarray}

Another measure of the approach of the density distribution to
 the fully-mixed state
 is the mean square difference in the cell occupation numbers, or
\begin{eqnarray}
D_2(n_i,{n_0}_i) & = &\sum_i\left[{{n_i -
{n_0}_i}\over {n_0}_i}\right]^2 {n_0}_i \\
             & = & \sum_i n_i\left[{n_i\over{{n_0}_i}} 
-1\right].
\label{L2_norm}
\end{eqnarray}
Similar measures of distance were adopted by Mahon {\it et al.}
(1995) and by Merritt \& Fridman (1996). 
This distance measure is equivalent to the entropy-based measure
$D_1$ when $n_i \approx {n_0}_i$.
Once again, $D_2 =0 $ in the fully mixed state.

We found that $D_1$ and $D_2$ behaved in similar ways with time
but that $D_2$ was more noisy; hence we present plots only of $D_1$ below.

In a real galaxy as in our ensembles, the total number of stars is finite.
Consequently, even in a fully-mixed state, an arbitrarily selected
volume of phase space (or configuration space) may not
contain a single star even if the probability of finding a star
in that volume is high. 
This provides us with a physical constraint on
the degree of coarse graining to apply.
Too coarse a mesh would show little evolution after early times 
and too fine a mesh would show large fluctuations even at late times. 
We defined our mesh as a $10\times 10 \times 10$ cubic grid with
edge length slightly greater than the amplitude of the axial 
orbit.
We found from experimenting with larger and smaller meshes that the 
errors arising from large fluctuations in cell occupancy were 
generally lower for this mesh.
When computing $\rho_{MC}$, densities lower than
that corresponding to one particle in a grid cell were assigned the
value zero to avoid comparing the cell densitites at the outer
edges of the distribution with infinitesmally small values.

\subsection{Approach to a Fully-Mixed State}

We evolved a number of isoenergetic ensembles of $10^4$ particles in 
two of our model potentials: $m_0=10^{-3}, M_{BH}=0$ (hereafter 
Model 1), and $m_0=10^{-1}, M_{BH}=3\times 10^{-3}$ (Model 2).
The starting points of the particles in each ensemble were chosen 
randomly from a small patch on the equipotential surface, 
surrounding one of the points on the regular grid of initial 
conditions, as shown in Fig. 7.
All ensembles had an energy equal to that of the $x$-axis orbit 
that just touches the ellipsoidal shell dividing the model into 
two equal-mass parts.
Each patch was approximately square and
had an edge length roughly equal to the separation between grid points.
Two of the ensembles from each model were centered about the 
starting points for trapped stochastics orbit, while the 
remainder were centered about non-trapped stochastic orbits.
The evolution time was $200 T_D$ for all ensembles -- an interval
that is comparable to the dynamical age of most of the stars in
a bright elliptical galaxy.

We chose the starting points to lie on the equipotential surface 
because only then could we be reasonably certain that most of 
orbits in an ensemble were stochastic.
However it is still possible that some of the particles in each ensemble 
lay close to a stable, high-order resonance and thus were regular.

The evolution of one ensemble, No. 2 from Model 1, is 
illustrated in Figure 8.
The starting points for this ensemble surround a strongly 
stochastic orbit and the mixing takes place very rapidly. 
After 15 dynamical times (roughly 5 Liapunov times) 
the clump has evolved into a filament containing several knots,
although some particles have broken free of the filament.
After about 30 dynamical times, the ensemble appears to have 
filled most of the volume beneath the equipotential surface, 
though in a highly nonuniform way.
By about 75 dynamical times, the ensemble appears to be 
approaching a time-invariant density distribution.
Most of the particles in this near-invariant distribution 
fill a region similar in shape to that of a regular box orbit.

The distribution of points at the final time step, $T=200 T_D$, 
is shown in Figure 9 for two ensembles.
The distributions appear to be nearly symmetric and unevolving 
at this late time; the snapshots of Figure 9 have accordingly 
been symmetrized about the principal planes to improve the 
statistics.
The density is highest in both cases near the $x$ axis and lowest 
near the $z$ axis -- in other words, the distribution has the 
same approximate shape as the model density.

The density distributions of Figure 9 represent nearly unchanging 
populations of chaotic phase space.
As such, they constitute bona fide building blocks for galaxies.
We assume that nature would make use of these invariant densities
in much the same way that it makes use of regular orbits.

The rate at which the different ensembles fill the 
volume within the equipotential surface, as measured by our
parameter $F$, is shown in Figure 10.
We see that the non-trapped ensembles evolve to fill most or all 
of this volume after only a few tens of crossing times, i.e.
of order ten Liapunov times.
The filling rate is highest for ensembles whose starting points 
lie near the $z$-axis or near the $y-z$ plane -- the part 
of initial-condition space which corresponds to the highest Liapunov 
exponents.
The trapped orbits, on the other hand, fill only a fraction of the
accessible volume even after $200$ dynamical times.
The value of $F$ oscillates strongly for these trapped ensembles as
the phase points slosh coherently from one side of the 
potential to the other -- the near-quasi-periodicity of the 
trajectories causes them to remain correlated for a long 
time.
(We carried out some additional experiments with ensembles in regular 
phase space.
They maintained their coherence even longer than the trapped 
stochastic ensembles, mixing almost not at all during the 200 
dynamical times.)
We note also that there is a continuity in mixing behaviors between the 
``trapped'' and ``non-trapped'' ensembles.
For instance, ensemble No. 2 from Model 1 (non-trapped) evolves
about as quickly as ensemble No. 6 (trapped).

A better measure of the departure from a fully-mixed state is 
shown in Figure 11, which plots $D_1(n_i,{n_{MC}}_i)$, the 
distance from the micro-canonical density, as a function of time.
These curves are very similar to those presented by Kandrup \& 
Mahon (1994) in their study of mixing in a two-dimensional, 
truncated Toda lattice, and many of the points made by them apply 
to our results as well.
The decay of $D_1$ is initially roughly exponential; the 
convergence rate $|d\ln D_1/dt|$ is approximately 
$0.03-0.1 T_D^{-1}$ for the non-trapped ensembles in both models, 
but rather longer (and not so clearly exponential) for the 
trapped ensembles.
This decay eventually slows, at which point $D_1$ for the non-trapped 
ensembles is fairly small, $0.2<D_1<0.5$, corresponding to 
a nearly-uniform filling of the energy surface.
Of course a nonzero, asymptotic value for $D_1$ is expected 
since the fully-mixed density is different from the 
micro-canonical density.
However the fact that different ensembles tend to significantly 
different values of $D_1$ over these timescales indicates that
the slow-down in the evolution of $D_1$ is due to a real decrease 
in the rate of mixing.

\subsection {Effect of Noise on the Mixing Rate}

Smooth, symmetric potentials are idealized representations of real 
galaxies, which often exhibit deviations from ellipsoidal 
symmetry, imbedded disks, or other fine structure.
As seen by a single star, these distortions would add
small-amplitude perturbing forces to the 
smooth forces produced by the overall mass distribution.
There may be other sources of time-dependent forces as well, 
including close encounters with stars in the dense central cusp, tides 
from satellite galaxies, etc.
During galaxy formation such perturbations would be 
very large, and slow oscillations in the potential might persist 
thereafter for many dynamical times as a galaxy evolves 
toward a steady state.

These weak or slow perturbations would not be expected 
to have a big effect on either the regular, or the strongly stochastic,
orbits.
The former would preserve their adiabatic invariants, while the 
latter move nearly randomly through phase space whether or 
not they are perturbed.
But small perturbations might have an appreciable effect on the 
mixing rate of weakly stochastic or trapped orbits, since the 
perturbations could carry stars away from a trapped region into
a more strongly stochastic region.
Goodman \& Schwarzschild (1981) found that perturbations had 
relatively little effect on the behavior of orbits in their 
Hubble-law potential; however Habib, Kandrup \& Mahon (1995), in a study of 
motion in a two-dimensional potential, found that even small 
amounts of noise could greatly enhance the mixing.

Following Goodman \& Schwarzschild (1981), we introduced noise 
into our mixing calculations by adding random impulses that changed 
the instantaneous direction of the velocity vector of a particle at
regular intervals.
All of the particles in the ensemble were perturbed together, 
although no two particles received the same perturbation.
This scheme left the energies of the particles unchanged.
Habib, Kandrup \& Mahon (1995) employed a more sophisticated prescription 
for adding noise but their scheme was also designed to preserve 
energy; thus our results might be roughly comparable to theirs.

The average rate of change of the transverse velocity due to 
perturbations is
\begin{equation}
\delta_t v_{\perp}^2 = Dv_{\perp}^2/Dt.
\end{equation}
A physically interesting number with which to compare this is
the standard diffusion coefficient for star-star encounters,
\begin{equation}
\langle{(\Delta v_{\perp})^2}\rangle =
8\pi G^2 m_f^2 n_f {3\over 2}{{\ln(0.4N)}\over{v_{mf}}} \times 0.583,
\label{diffusion}
\end{equation}
(Spitzer \& Hart 1971), where $v_{mf}$, the rms velocity of the 
field stars, has been assumed to be equal to that of the test star.
We can estimate $\langle{(\Delta v_{\perp})^2}\rangle$
for our models by assuming that the total number of stars in the 
system is $10^{11}$ each with a mass of $1 \Msolar$, and 
identifying $n_f$ with ${\overline \rho} /m_f$, where $\overline \rho$
is the mean density within the half-mass radius. 
The mean velocity of the field stars is taken to be $v_{mf} = 4 X_0/T_0$ 
where $X_0$ and $T_0$ are the amplitude and period of the $x$-axial 
orbit with the same energy as that of our ensembles.
We find $\langle{(\Delta v_{\perp})^2}\rangle \approx 2\times 10^{-9}$ 
in model units, corresponding to a relaxation time of $\sim 10^7 T_D$.

We then define the strength of the imposed perturbations via the 
parameter $\eta$, where
\begin{equation}
\eta = \delta_t v_{\perp}^2 / \langle{(\Delta 
v_{\perp})^2}\rangle.
\end{equation}
We chose $\eta$ to have one of the three values $1, 10^2$ or 
$10^4$, similar to the choices made by Goodman \& Schwarzschild 
(1981).
The last choice is clearly an overestimate of the perturbation size 
to be expected from star-star encounters, but it is still quite 
modest when interpreted as a representation of perturbations from 
other possible sources.
The two-body relaxation time as we have defined it is of order 
$10^7$ dynamical times in our models, so a value of $\eta=10^4$ 
corresponds to a perturbation timescale of $1000 T_D$ -- quite a 
weak perturbation.

While in principle the time interval
between successive perturbations should be described by a Poisson
distribution and the amplitude of the individual perturbations should be
random, we have assumed for simplicity of programming that the
perturbations occur at regular intervals and are of equal amplitude.
Having specified $\eta$, one still has the freedom to specify how 
often the perturbations are applied.
We followed the example of Goodman \& Schwarzschild (1981) by 
perturbing the particles roughly once per dynamical time.

The results are shown in Figure 12, for $\eta=10^4$;
smaller amplitude perturbations were found not to significantly affect 
the mixing rate of any ensemble.
We find that noise with this amplitude is quite effective 
at accelerating the mixing.
All of the ensembles now evolve in a similar way, and at late 
times, every ensemble appears to be approaching a common final state.
Furthermore, the distinction between trapped and non-trapped ensembles
is much reduced in the presence of noise.
It seems reasonable to conclude that random perturbations 
can significantly enhance the mixing rates of ensembles of stochastic
orbits, causing even ensembles of trapped orbits to reach an invariant
distribution in perhaps a few hundred orbital times.

\subsection{Mixing Timescales}

We see in Figures 11 and 12 evidence of two timescales associated 
with the mixing.
The mixing is initially fast, with a roughly exponential 
dependence of $D_1$ on time.
Subsequent mixing is slower and continues until the end of the 
integrations at $t=200T_D$.
These two regimes are evident in all of the integrations that 
included noise.
In the absence of noise, the non-trapped ensembles evolve in this 
way but the evolution of the trapped ensembles is more complex.

The existence of a second, longer timescale for the orbital 
evolution was noted above in the discussion of the Liapunov exponents.
It is reasonable to suppose that the slower mixing
seen in Figures 11 and 12 takes place on roughly the same timescale over
which the Liapunov exponents were observed to evolve toward a 
common value in \S3 -- i.e. hundreds or thousands of 
orbital times.
After elapsed times of this order (or perhaps less in the integrations 
including noise) we would expect all of the stochastic ensembles 
to have attained very nearly the same density distribution. 
Limited computing resources kept us from verifying this prediction.

Kandrup and coworkers (Kandrup \& Mahon 1994; Mahon et al. 1995; 
Habib, Kandrup \& Mahon 1995) also noted the existence of a long 
and a short timescale associated with mixing in their two-dimensional 
potentials.
Kandrup \& Mahon (1994) found the initial mixing rate to be 
roughly 0.15 times the rate of divergence of nearby trajectories 
in their truncated, Toda-lattice potential.
The initial rate of decay of $D_1$ in our non-trapped 
ensembles is difficult to estimate with any accuracy.
At first the decay is very rapid, and appears to take place at a 
rate that is approximately equal to the inverse Liapunov time.
The decay rate continually slows, however, and an average rate over the 
first 50 orbital times is 
of order $0.03-0.1 T_D^{-1}$ in the absence of noise, 
compared to a typical Liapunov exponent of $0.3 T_D^{-1}$.
The corresponding ratio is $\sim 0.1-0.3$, in good agreement 
with Kandrup \& Mahon's result.
This agreement is not surprising: it simply reflects the fact 
that the mixing is driven initially by the orbital instability
and so its characteristic time is tied to the Liapunov time.
We assume that mixing would continue at this rate except for 
the existence of barriers in phase space, i.e. invariant tori and 
``cantori'' that inhibit the diffusion (Mackay, Meiss \& Percival 1984).

We note that certain features of the $D_1(t)$ curves are 
dependent on the details of our numerical treatment.
If our initial conditions had been selected from smaller patches 
on the equipotential surface, or if we had used a finer grid to 
compute $D_1$, we would have observed a longer exponential decay, 
because more time would have been required for the 
mixing to eliminate correlations on the scale of the grid cells. 
Stated differently, the coarse-grained relaxation time depends 
on the level of coarse-graining.

However, the rate $|d\ln D_1/dt|\equiv T_M^{-1}$ at which correlations initially 
decay should not be a strong function of these details.
Kandrup \& Mahon (1994) found this to be the case, as did we, 
based on a limited number of additional experiments with 
different numbers of grid cells.

We did not investigate the dependence of the mixing rate on 
energy. 
However Kandrup \& Mahon (1994) found a fairly constant ratio 
between the Liapunov and mixing timescales at different energies 
in their two-dimensional potentials, and 
the same may be true in our models.
Merritt \& Fridman's (1996) results (their Figure 8b) suggest that 
this is approximately true.

Which of these two timescales should we associate with the mixing 
in real galaxies?
The answer is probably: both.
The initial, exponential decay of $D_1$ reflects the erasure of 
correlations due to the instability of the motion; after several 
of these decay times have elapsed, the ensemble has filled, in a 
coarse-grained sense, most of the configuration-space 
region accessible to it.
Inspection of Figure 8 and other plots like it suggests that a 
time of $\sim 100 T_D$, i.e. $\sim5T_M$, can be roughly identified 
with this ``relaxation time.''
This is an astronomically interesting timescale; orbital 
periods at the half-light radii of bright elliptical galaxies are 
of order $10^{-2}$ times the age of the universe (\S6).

The slower mixing that is associated with diffusion into the Arnold 
web might cause a galaxy to continue to evolve in shape
as the stochastic orbits mix toward their invariant distributions.
We were not able to estimate this longer timescale with any precision, 
but there are a number of indications that it is of 
order $10^3 T_D$.
Such a number is consistent with the slow decay seen on Figures 
11 and 12, and with rate of approach of the Liapunov exponents toward 
a common value in Figures 3-5.
Merritt \& Fridman (1996) also found that the time required for 
single stochastic orbits to behave ``ergodically'' was of order 
$10^3 T_D$.

These timescales might be substantial overestimates if galaxies 
are efficiently mixed during their formation.
This possibility is discussed further below.

In potentials with larger cores or smaller central singularities, 
we showed above that a larger fraction of the stochastic orbits 
mimic regular orbits for hundreds or thousands of oscillations.
We expect mixing to be less efficient in these models,
with the mixing rate approaching zero as $m_0\rightarrow 1$ or 
$M_{BH}\rightarrow 0$.
It follows that galaxies might differ enormously in the degree
to which their stochastic orbits are mixed, a point that we
return to in \S6.

\section {Discussion}

We discuss three issues raised by this work: the applicability of
Jeans's theorem to systems containing stochastic orbits; the relation
between chaotic mixing and violent relaxation; and the importance of
chaos for the structure and slow evolution of elliptical galaxies.

\subsection {Jeans's Theorem}

Jeans's theorem specifies the conditions under which a 
collisionless stellar system will be in a state of equilibrium.
In modern texts (as well as in Jeans's original formulation) 
the theorem is usually restricted to systems 
that are fully integrable, i.e. in which all or virtually all of 
the orbits respect a number of isolating integrals equal to or 
greater than the number of degrees of freedom.
Binney \& Tremaine (1987) call this restricted version 
the ``strong Jeans theorem'' and base its derivation 
on the time-averages theorem, i.e. the ergodic property of regular orbits.
Binney (1982b) argues further that the lack of a general proof of 
ergodicity for irregular orbits calls into question the
applicability of Jeans's theorem to more general, non-integrable 
systems.

We do not see a clear link between ergodicity -- which is a time-averaged 
property of trajectories -- and stationarity, which is a 
statement about conditions at a single time.
We would accordingly state the requirements for a stationary state
in a different way, even in systems that are fully regular.
Suppose one divides the phase space of a system with a fixed potential
into a set of regions, each of which is filled by a single orbit.
For regular orbits, these regions are invariant tori, while for 
irregular orbits the regions are more complex in shape.
A sufficient condition for a steady state is that the phase-space 
density $f$ be constant within each of these regions.
The proof follows directly from Liouville's theorem: since $f$
is conserved following the flow, an initially constant value 
within any bounded region will remain constant forever.
This proof relies purely on the incompressibility of 
the flow; no additional properties of the motion -- such as 
ergodicity or mixing -- are required.\footnote[3]{Just this point 
of view is implicitly taken by anyone who writes $f=f(E,L_z)$ for 
an axisymmetric galaxy.
Such a distribution function assigns nonzero densities to 
stochastic parts of phase space, in general, and yet it describes 
a steady state since $f$ is forced
to be constant throughout any chaotic subregions of the hypersurfaces 
defined by constant $E$ and $L_z$.
The ergodicity of the motion in those stochastic regions is of no
importance.}
We therefore disagree with derivations of Jeans's theorem
that imply a link between stationarity and ergodicity.

The situation is not quite this simple, however, since
collisionless stellar systems have no built-in mechanism by which 
the fine-grained distribution function can evolve to such a 
state.
The importance of properties like ergodicity or mixing -- which 
are effectively measures of the degree of randomness of the flow 
(Lichtenberg \& Lieberman 1983, p. 273) -- are that they can 
sometimes guarantee the approach to a coarse-grained steady state.
The way this works in the case of a purely regular
system was discussed in \S1, and the evolution of chaotic 
flows toward near-invariant distributions was the subject of \S5.
Thus, while the possibility of stationary configurations depends only on 
Liouville's theorem, which is valid generally, relaxation
toward a coarse-grained steady state will only occur if the flow 
satisfies more stringent conditions.

Another relevant point is the difficulty of dividing phase space 
into ``chaotic'' and ``regular'' regions.
The Arnold web is believed to permeate the entire phase space, intersecting 
or lying infinitesimally close to every point, even points 
associated with regular orbits (Lichtenberg \& Lieberman 1983, p. 54).
It would therefore be extremely difficult in practice to define what is 
meant by the ``chaotic part of phase space'' at a given energy.
However one could imagine approximating a uniformly-filled, chaotic 
phase space by evolving an ensemble of points until their 
coarse-grained density was nearly unchanging, as was done here in 
\S 5.
The required evolution time might be extremely long, although the addition 
of noise to the forces might accelerate the mixing.
Such a scheme would allow one to incorporate the chaotic parts of 
phase space into time-independent models.
Nature presumably does something very much like this when it 
makes real galaxies.

\subsection {Violent Relaxation}

In a real galaxy, however, the approach to a stationary state 
takes place against the background of a time-varying potential.
Following Lynden-Bell's (1967) pioneering paper, potential
fluctuations have generally been credited with producing the
relaxation.
Lynden-Bell identified the collisionless relaxation time as
\begin{equation}
T_r={3\over 4} \langle{\dot\Phi^2\over\Phi^2}\rangle ^{-1/2},
\end{equation}
with $\dot\Phi$ the rate of change of the gravitational 
potential, and argued that ``violent relaxation'' should 
take place on roughly this timescale.

It is clear that a rearrangement of stars in energy space 
can only take place if the gravitational potential is time-dependent.
But it is less clear that there exists a deeper link between 
potential fluctuations and collisionless relaxation.
Our experiments (and those of Kandrup and co-workers) 
demonstrate that a time-varying potential is not necessary for 
strong evolution of the phase-space distribution to take place: mixing 
toward a steady state can occur in a completely fixed potential.
(Phase mixing is of course another example of such evolution, as 
pointed out by Lynden-Bell.)
Nor is relaxation guaranteed to occur simply because the 
potential is fluctuating.
Consider for example a group of stars that sit motionless 
at the center of a collapsing proto-galaxy. 
The energy of these stars changes as the galaxy collapses, but
no relaxation of any physically interesting sort 
takes place: the stars remain fixed in phase space as their energies
change.
More generally, one can imagine enclosing a stellar system inside 
a massive spherical shell whose radius is varied with time.
The potential within the galaxy will fluctuate, but there 
are no forces from the shell and hence no relaxation.\footnote[4]{We 
thank R. Miller for suggesting this example.}
Changes in the potential imply here only a relabeling 
of the particle energies; they do not, by themselves, 
constitute relaxation.

A time-varying potential can even {\it inhibit} evolution.
For instance, the stars in a dense satellite that spirals into a 
larger galaxy are slow to mix, because the self-gravity of the 
satellite maintains the coherence of their motion.
If this self-gravity were turned off, allowing the stars to move 
in the {\it fixed} potential of the larger galaxy, they would be free 
to phase-mix.
Thus the time-varying component of the potential acts in this 
case to maintain correlations, not to destroy them.

It seems evident that relaxation -- by which we mean ``evolution 
to a stationary state'' -- in collisionless stellar systems 
always requires a divergence in particle trajectories, that is to 
say mixing, 
since only through mixing can memory of the initial conditions be erased.
Potential fluctuations constitute relaxation only to the 
extent that they promote mixing.
Many of the specific physical mechanisms that are thought to cause
collisionless relaxation during galaxy formation can be usefully 
thought about in this way.
For instance, in ``potential scattering,'' a star gains energy by 
falling in through a deep potential well and escaping through a 
shallow one.
But this energy gain is of no importance by itself; it leads to 
relaxation only if nearby stars gain different amounts 
of energy, since otherwise those stars will remain in fixed positions 
relative to one another.
In other words, potential scattering contributes to relaxation only 
insofar as it causes initially nearby particles to move apart.
But it is the divergence in phase space, and not the 
energy change per se, that is correctly identified with the 
relaxation.

If the potential fluctuations are sufficiently rapid or strong,
we might guess that chaotic mixing -- or something like it -- 
will take place in no more than a collapse or crossing time, 
as in Lynden-Bell's picture.
The rapid approach to equilibrium seen in $N$-body experiments
suggests that this is approximately the case.
However it is still the mixing process, and not the potential 
fluctuations themselves, that is responsible for 
the relaxation.
The chaotic mixing experiments described here in a fixed 
potential might therefore 
provide a simple picture of the way in which collisionless relaxation 
takes place under more general conditions.

\subsection {Importance of Chaotic Mixing for the Structure and 
Evolution of Elliptical Galaxies}

Schwarzschild (1993) and Merritt \& Fridman (1996) argue that 
many or most of the stars in triaxial stellar systems could 
be on stochastic orbits.
If so, elliptical galaxies might evolve on timescales that
are similar to the timescales for chaotic mixing derived here.

One problem with this simple view is the difficulty of comparing 
the mixing rates in the regular and stochastic parts of phase 
space.
It is often loosely stated that regular orbits fill their tori 
in a short time -- some small multiple of an orbital period -- 
while stochastic orbits diffuse more slowly.
According to this view, any slow evolution would be driven purely 
by the diffusion of the stochastic orbits.
But such statements confuse ergodicity with mixing.
While the {\it time averaged} density of stars on an invariant 
torus approaches a coarse-grained steady state in a few crossing times, 
an {\it ensemble} of stars on a regular orbit never relaxes.
One could take the extreme view that mixing of stochastic orbits 
is infinitely faster than that of regular orbits, since a collection of
stars confined to a single regular orbit does not mix!

In fact it seems likely that the evolution to a near-stationary 
state takes place mostly during galaxy 
formation, when the potential is strongly time-dependent and 
the motion is essentially chaotic.
Once the potential begins to settle down, some stars will find 
themselves in regular parts of phase space and some in stochastic 
parts; in regions of both types, the phase space distribution will be 
highly mixed and, perhaps, not far from a coarse-grained steady state.
Subsequent evolution would be driven both by phase mixing and by 
chaotic mixing.
We do not know which of these two types of mixing would dominate 
the evolution in a typical case; perhaps both would be important.

However, the consequences of continued mixing would be very 
different for the two types of orbit.
A non-mixed region of regular phase space -- a non-uniformly
populated torus, for example -- would almost always generate a 
lopsided configuration-space density (Eq. 2).
But an ensemble of stochastic orbits can be symmetric in 
configuration space even if it does not fully populate its 
allowed phase space region.
In fact ensembles of trapped stochastic orbits appear to evolve 
in much this way, quickly reaching a nearly symmetric shape
before more slowly diffusing into the full Arnold web.
Real elliptical galaxies often -- though not always -- appear 
to have a high degree of symmetry, which suggests 
that their regular parts of phase space are nearly fully mixed.

There is a second way in which the consequences of 
phase mixing and chaotic mixing are different.
Mixing in stochastic phase space leads to a reduction in the 
effective number of orbits, since it replaces all the 
stochastic trajectories at a given energy by a single invariant 
density distribution.
If chaotic mixing timescales are short compared to a galaxy 
lifetime, this reduction might 
force a galaxy to evolve away from a triaxial shape toward an 
axisymmetric one (Schwarzschild 1981).
On the other hand, if chaotic mixing timescales are long, then elliptical
galaxies might persist in slowly evolving, approximately 
triaxial shapes.
Which of these two descriptions is most correct depends only on 
the ratio of the chaotic mixing timescale to the galaxy lifetime
(and, of course, on the fraction of stars that follow chaotic 
trajectories).

The results presented here and by other workers suggest that 
this ratio is strongly dependent on the central structure of a galaxy.
In a triaxial model with a low central concentration of mass, 
many of the boxlike orbits may be stochastic, but these 
stochastic orbits behave very much like regular orbits for hundreds or 
thousands of crossing times.
Goodman \& Schwarzschild (1981) first described such behavior in 
their study of a triaxial model with a Hubble-law profile, and we 
observed similar behavior for the orbits in our models with large $m_0$ and 
small $M_{BH}$.
Models with cores do not describe early-type galaxies very well.
However Merritt \& Fridman (1996) found that the stochastic 
motion in triaxial models with shallow cusps, $\rho\propto 
r^{-1}$, was similar to that seen here in models with cores.
This is reasonable since the gravitational force generated by 
a weak power-law cusp is finite near the center.

On the other hand, a steep central cusp or a massive central 
singularity can persuade most of the stochastic orbits to behave 
ergodically over much shorter timescales.
Precisely how steep a cusp is required is not yet known.
The models studied here have a fixed central slope, $\rho\propto 
r^{-2}$, for small $m_0$ -- the same central dependence as in
Schwarzschild's (1993) scale-free models and in Merritt \& 
Fridman's (1996) ``strong cusp'' model.
Presumably the transition from effectively regular behavior to 
effectively ergodic behavior for the majority of stochastic 
orbits takes place somewhere between $\rho\propto r^{-1}$ and 
$\rho\propto r^{-2}$; one would like to know exactly where.
In the case of central singularities, the results of \S 4 
suggest that the critical mass for inducing ergodic behavior 
is about $0.3\%$ of the galaxy mass.

The strong dependence of the orbital behavior on the degree of 
central mass concentration suggests that triaxial galaxies might 
come in two distinct families: those in which the orbital motion is 
nearly regular, and those in which the stochastic parts 
of phase space are nearly mixed.
The latter galaxies might not persist for long in triaxial configurations 
due to the limited number of orbital shapes that would be available for 
maintaining a non-axisymmetric shape.

One consequence is that faint elliptical galaxies are less
likely to be triaxial than bright ones.
Ellipticals with absolute luminosities less than $M_B\approx-20$
have the steepest central density cusps on average, with logarithmic 
slopes $\gamma$ in the range $-2\lap\gamma\lap-1$, compared to 
$0\lap\gamma\lap-1$ for brighter ellipticals (Gebhardt et al. 1996).
Low-luminosity ellipticals also have higher average densities,
and shorter dynamical times, than bright ellipticals; all 
else being equal, the mixing will have gone farther in these galaxies.

For an estimate of the dynamical ages of bright ellipticals, we refer
to Katz \& Richstone's (1985) study of three galaxies, NGC 4472 
($M_B=-22.1$), 4374 ($-21.1$), and 4636 ($-20.9$).
Adopting their inferred $M/L$ of 19.5 in solar units ($H_0=75$), 
we find for the circular orbital period at the effective radius 
$1.3\times 10^7$, $4.3\times 10^7$ and $1.86\times 10^7$ yr, 
respectively.
Assuming a galaxy lifetime $T$ of $5\times 10^9$ yr, the 
``dynamical ages'' $T/T_D$ of these bright galaxies are 
$38$, $115$ and $27$ at the half-light radius.
We estimated above (\S5) that relaxtion to a strongly mixed state 
might require of order $10^2 T_D$ in a triaxial galaxy with a 
strong central concentration of mass.
Thus -- even assuming that these bright ellipticals have massive 
nuclear black holes -- we would expect them to be strongly mixed
only in their central regions, $r\lap r_e$.

For comparison, we computed dynamical ages of several 
low-luminosity ellipticals ($M_B\approx -19$) with known 
distances, assuming the same $M/L$ as above.
Typical half-mass orbital times were found to be 
$4\times 10^7$ yr, with dynamical ages in the range $100-200$.
The combination of larger dynamical ages with steeper central cusps 
implies that the stochastic orbits in low-luminosity, triaxial 
ellipticals would be more strongly mixed than in bright ellipticals.

(Bright ellipticals are also more likely to have undergone a
recent merger event, and in this sense too might be dynamically 
``younger'' than faint ellipticals.)

An extreme example is the nearby dwarf elliptical M32,
which is known to have a steep central density cusp, 
$\rho\propto r^{-1.63}$ (Gebhardt et al. 1996), as well as 
a central mass concentration containing $\sim 0.3\%$ of the stellar mass 
(Tonry 1987).
M32 is also atypically dense: the circular orbital time 
at the half-light radius is only about $10^7$ yr (Dehnen 1996),
implying that most of the stars in M32 have completed 
$10^3$ orbits or more since the galaxy formed.
Triaxiality would presumably be difficult to maintain in this galaxy, 
and in fact detailed modelling (Dehnen 1995; van der Marel et al 1994) 
suggests that M32 is nearly oblate.

There are two complications to this simple picture.
First, it is possible that most or all early-type galaxies contain 
massive central singularities (Kormendy \& Richstone 1995). 
If so, the steepness of the stellar cusp would be of secondary 
importance for determining the degree of stochasticity -- 
because the motion of the boxlike orbits would be strongly chaotic 
even in the absence of a cusp.
Second, faint elliptical galaxies are often rapidly rotating 
(Davies et al. 1983).
The effect of figure rotation on the degree of stochasticity in 
triaxial models has not yet been well explored.

Leaving aside for the moment these complications, we can ask 
whether there is any evidence that bright elliptical galaxies 
have systematically different shapes than faint ellipticals.
There is in fact an emerging concensus for such a dichotomy
(e.g. Kormendy \& Bender 1996): low-luminosity ellipticals are 
said to be ``disky'' and oblate, while high-luminosity 
ellipticals are ``boxy'' and triaxial.

The separation of elliptical galaxies into boxy and disky 
subgroups seems well established, but in our view the evidence 
that faint ellipticals are oblate while bright ellipticals are 
triaxial is not yet compelling.
The most convincing test for triaxiality is minor-axis rotation.
Franx, Illingworth \& de Zeeuw (1991) found that relatively few 
galaxies exhibit kinematic misalignments -- only about 8 galaxies 
in their sample of 38.
However only one of their galaxies was fainter than $M_B=-20$, 
and there is no obvious tendency in their sample for the degree of 
kinematic misalignment to correlate with luminosity.
Thus, little appears to be known at the present time about the relative 
frequency of minor-axis rotation in bright and faint ellipticals.

Tremblay \& Merritt (1996) found that the distribution of Hubble 
types for bright ellipticals ($M_B\lap -20$) was significantly different 
than for faint ellipticals: bright galaxies appear rounder
on average, and they have a distribution of apparent shapes 
that is difficult to reproduce under the axisymmetric hypothesis.
However the Hubble-type data are equally consistent with triaxial 
shapes for both bright and faint galaxies.

Detailed modelling of individual elliptical galaxies has only 
begun; the example of M32, which is faint and apparently oblate,
was discussed above.

We conclude that the observational data are consistent with a picture 
in which bright ellipticals are triaxial and faint ones oblate, 
but that such a picture is not yet compelling.

Within individual galaxies, the dynamical age is a function of 
radius, and we would sometimes expect the central regions of a triaxial
galaxy to be highly mixed while the outer regions are still in the 
process of mixing.
The transition between the two regions would occur at the 
``mixing radius'' $R_M$ where the mixing time is equal to 
the age of the galaxy.
Although it is difficult to calculate $R_M$ with any 
precision, we can obtain a crude estimate by equating the galaxy lifetime 
to the time $\sim 100T_D$ that was found above to characterize 
mixing in strongly stochastic potentials.
We find $R_M\lap r_e$ in the three bright galaxies discussed above 
and $R_M \gap r_e$ in the fainter ellipticals; 
while in M32 this radius would lie well outside of $r_e$
-- this galaxy should be strongly mixed throughout.
Even at radii less than $R_M$,
mixing would be expected to continue at the slower rate seen in 
the experiments of \S 5.

Mixing near the center of a triaxial galaxy would tend to 
make the galaxy axisymmetric there.
Observed from a random angle, such a galaxy would probably exhibit a 
non-constant ellipticity, and a major-axis position angle
that varied with radius.
Could this be the origin of isophote twists?

\bigskip

We were motivated to think about the issues discussed here 
in part by J. Binney's provocative paper on the 
applicability of Jeans's theorem to nonintegrable systems.
Discussions with him, and also with O. Gerhard, D. Heggie, H. Kandrup, 
J. Laskar, R. Miller, T. Padmanabhan, D. Pfenniger and S. Tremaine 
helped to sharpen our thinking.
W. Dehnen read the manuscript especially carefully and made a number
of comments that led to improvements in the presentation.
The computer programs used in this study for integrating orbits 
and computing Liapunov exponents were written by the Geneva group 
and graciously lent to us by S. Udry and D. Pfenniger.
B. Tremblay supplied some of the numbers in \S6.
This work was supported by NSF grant AST 90-16515 and by NASA 
grant NAG 5-2803 to DM.

\clearpage

\begin{table}
\caption{Liapunov Exponents}
\begin{tabular}{ccccc}
\\ \hline
$M_{BH}$ & $m_0$ & $\sigma_1T_D$ & $\sigma_2T_D$ & N \\ \hline
$0$ & $10^{-1}$ & $0.14\pm 0.06$ & $0.045\pm 0.02$ & 85  \\
    & $10^{-2}$ & $0.21\pm 0.09$ & $0.078\pm 0.03$ & 141 \\
    & $10^{-3}$ & $0.27\pm 0.05$ & $0.085\pm 0.02$ & 152 \\ \\

$10^{-3}$ & $10^{-1}$ & $0.15\pm 0.03$ & $0.066\pm 0.02$ & 163 \\
          & $10^{-3}$ & $0.27\pm 0.06$ & $0.080\pm 0.02$ & 151 \\ \\

$3\times 10^{-3}$ & $10^{-1}$ & $0.20\pm 0.04$ & $0.097\pm 0.02$ & 159 \\
                  & $10^{-3}$ & $0.28\pm 0.07$ & $0.086\pm 0.02$ & 152 \\ \\

$10^{-2}$ & $10^{-1}$ & $0.28\pm 0.08$ & $0.16\pm 0.04$ & 171 \\
          & $10^{-3}$ & $0.32\pm 0.10$ & $0.13\pm 0.04$ & 163 \\ \hline

\end{tabular}
\end{table}

\clearpage

\appendix
\section {APPENDIX}
\centerline{Computation of the Gravitational Forces}

Here we present expressions for the gravitational forces corresponding
to the triaxial mass models described in \S2, and their derivatives,
which are required when computing the Liapunov exponents.
We also discuss the numerical techniques used to evaluate the forces
quickly and accurately.

 The forces may be derived from the potential gradient by computing
 the contributions to the force separately from the two different
ellipsoidal coordinate systems. Thus if the potential is expressed as
${\bf \Phi(x) = \Phi^A(x) + \Phi^B(x)}$, then the forces are
 ${\bf F(x) = F^A(x) + F^B(x)}$, with
\begin{eqnarray}
 {\bf F^A}_x & = & - {{\partial{\bf \Phi^A}}\over{\partial x}},\\
 {\bf F^B}_x & = &- {{\partial{\bf \Phi^B}}\over{\partial x}}.
\end{eqnarray}
The required gradients can be evaluated by using
\begin{eqnarray}
 {{\partial}\over{\partial x}} & = &
2x\left[{{(\lambda-b^2)(\lambda-c^2)\, \partial}\over
{(\lambda-\mu)(\lambda-\nu)\, \partial \lambda}}
+{{(\mu-b^2)(\mu-c^2)\, \partial}\over{(\mu-\nu)
(\mu-\lambda)\,\partial \mu}}
+{{(\nu-b^2)(\nu-c^2)\, \partial}\over
{(\nu-\lambda)(\nu-\mu)\,\partial \nu}}\right].
\end{eqnarray}

The derivates with respect to the cartesian coordinates $ y, z$ are
similarly obtained by the substitutions
$$
\lambda  \rightarrow \mu  \rightarrow \nu \rightarrow \lambda,
\qquad               x    \rightarrow y  \rightarrow z,
\qquad      a  \rightarrow  b  \rightarrow   c  \rightarrow a.
$$

It can  be shown that
\begin{eqnarray}
{{\partial G(\tau)}\over {\partial \tau}} & = & { 2\over
3}R_J(a^2,b^2,c^2,\tau), 
\end{eqnarray}
where $R_J(a^2,b^2,c^2,\tau)$
is Carlson's symmetrized version of the incomplete elliptic
integral of the third kind which can be numerically evaluated by fast
algorithms.

The forces are then given by
\begin{eqnarray}
{\bf F}_x^A & = &{{2x}\over{3\pi(1-m_0)}} \left[Q^A_\lambda
R_J(a^2,b^2,c^2,\lambda)
+ Q^A_\mu R_J(a^2,b^2,c^2,\mu)
+ Q^A_\nu R_J(a^2,b^2,c^2,\nu)
\right] \\
{\bf F}_x^B & = & {\frac {-2x }{3\pi(1-m_0){m_0}^2}}[
Q^B_{\lambda^\prime} R_J(a^2,b^2,c^2,\lambda^\prime/m_0^2) 
 + Q^B_{\mu^\prime} R_J(a^2,b^2,c^2,\mu^\prime/m_0^2) \nonumber\\
\quad & & + Q^B_{\nu^\prime} R_J(a^2,b^2,c^2,\nu^\prime/m_0^2)]
\end{eqnarray}
where we have adopted the shorthand notation
\begin{eqnarray}
Q_\tau^A & = &
{\frac{(\tau-b^2)(\tau-c^2)}{(\tau-\tau_{+})(\tau-\tau_{-})}},\nonumber\\
\label{qst}
S_\tau^A & = &
{\frac{(\tau-c^2)(\tau-a^2)}{(\tau-\tau_{+})(\tau-\tau_{-})}},\\
T_\tau^A & = &
{\frac{(\tau-a^2)(\tau-b^2)}{(\tau-\tau_{+})(\tau-\tau_{-})}}, \nonumber
\end{eqnarray}
where since  $$\lambda \rightarrow \mu \rightarrow \nu \rightarrow \lambda$$
$\tau \equiv \lambda$ implies  $\tau_{+} \equiv
\mu$  and $\tau_{-} \equiv \nu$  and likewise
for the second set of coordinates ($\lambda^\prime, \mu^\prime,
\nu^\prime$) with $a^2, b^2, c^2$ replaced by $a^2m_0^2, b^2m_0^2, c^2m_0^2$
respectively.
Similar expressions are obtained for the $y$ and $z$ components.
This form of the force equations entails evaluation of the elliptic
integral $R_J$ for which fast algorithms exist.

The force equations in this simplified notation are then given by
\begin{eqnarray}
{\bf F}_x & = &{\frac{2x}{3\pi(1-m_0)}}\left[\sum_\tau Q_\tau^A
R_J(a^2,b^2,c^2,\tau) - {\frac 1 {m_0^2}}\sum_{\tau^\prime}
Q_{\tau^\prime}^BR_J(a^2,b^2,c^2,{\tau^\prime\over{m_0^2}})\right],
\nonumber \\
{\bf F}_y  & = & {\frac{2x}{3\pi(1-m_0)}}\left[\sum_\tau S_\tau^A
R_J(a^2,b^2,c^2,\tau) - {\frac 1 {m_0^2}}\sum_{\tau^\prime}
 S_{\tau^\prime}^B
R_J(a^2,b^2,c^2,{\tau^\prime\over{m_0^2}})\right],
 \\
{\bf F}_z & = & {\frac{2x}{3\pi(1-m_0)}}\left[\sum_\tau T_\tau^A
R_J(a^2,b^2,c^2,\tau) - {\frac 1 {m_0^2}}\sum_{\tau^\prime}
T_{\tau^\prime}^BR_J(a^2,b^2,c^2,{\tau^\prime\over{m_0^2}})
\right].
\nonumber
\end{eqnarray}

The derivatives of the forces are obtained by differentiating the 
above expressions for the force.
We have
\begin{eqnarray}
{\frac{\partial R_J(a^2,b^2,c^2,\tau)}{\partial \tau}}
& = & -{\frac 3 2}{\int_0^\infty \frac{du}{(u+\tau)^2
{\sqrt{(u+a^2)(
u+b^2)(u+c^2)}}}}.
\end{eqnarray}
We  adopt the notation
\begin{eqnarray}
\tilde{R}_J(\tau) & = & {\frac{\partial R_J(a^2,b^2,c^2,\tau)}{\partial 
\tau}},
\end{eqnarray}
with the understanding that $\tau$ represents either the first set of
coordinates or $\tau^\prime/m_0^2$.
These improper integrals can be rewritten as proper integrals by a change
of variable, $s^2 = a^2/(u+a^2)$, so that
\begin{eqnarray}
\tilde{R}_J(\tau) & = &{-3\over {a^5}}\int_0^1{\frac{s^4 ds}
{(1-\hat{\tau}s^2)^2
\sqrt{(1-\hat{b^2}s^2)(1-\hat{c^2}s^2)}}},
\end{eqnarray}
where
$$\hat{\tau} = 1-{\frac {\tau}{ a^2}},\qquad \hat{b^2} =
1-{\frac {b^2} {a^2}},
\qquad \hat{c^2} = 1-{\frac {c^2} {a^2}},$$
and similarly for $\tilde{R}_J(\tau^\prime/m_0^2)$.

Also,
\begin{eqnarray}
 {\frac {\partial Q_\tau^A}{\partial \tau}} & = &{\frac\partial{\partial
\tau}}
\left[{\frac {(\tau -b^2)(\tau-c^2)}{(\tau
-\tau_{+})(\tau-\tau{-})}}\right]\nonumber\\
\quad & = & {\frac{(\tau -b^2)(\tau-c^2)}{(\tau -\tau_{+})(\tau-\tau{-})}}
\left[{1\over{(\tau-b^2)}}+{1\over{(\tau-c^2)}}-{1\over{(\tau-\tau_{+})}}
-{1\over{(\tau-\tau_{-})}}\right] \\
\quad &  \equiv & Q_\tau^A \tilde{Q}_\tau^A \nonumber\\
{\frac {\partial Q_{\tau_+}^A}{\partial \tau}}  & = &
{\frac\partial{\partial \tau}}
\left[{\frac {(\tau_{+} -b^2)(\tau_{+}-c^2)}{(\tau_{+} -\tau)
(\tau_{+}-\tau_{-})}}\right] \nonumber \\
\quad & = &  {\frac{Q_{\tau_+}^A}{(\tau_{+}-\tau)}}.
\end{eqnarray}
Similarly,
\begin{eqnarray}
{\frac {\partial Q_{\tau^\prime}^B}{\partial \tau^\prime}} & = &
Q_{\tau^\prime}^B \tilde{Q}_{\tau^\prime}^B, \\
{\frac {\partial Q_{\tau_+^\prime}^B}{\partial \tau^\prime}} & = &
{\frac{Q_{\tau_+^\prime}^B} {(\tau_{+}^\prime-\tau^\prime)}}.
\end{eqnarray}

With the above notation the force derivatives can be written
as
\begin{eqnarray}
{\bf F}_{xx} & = & {\frac{{\bf F}_x}{x}} + {\bf F^A}_{xx} + {\bf
F^B}_{xx}
\end{eqnarray}
where
\begin{eqnarray}
{\bf F^A}_{xx} & = & {\frac{4x^2}{3\pi(1-m_0)}}[
\sum_\tau{({{Q_\tau}^A})^2 \tilde{Q}_\tau^A R_J(\tau)} -
\sum_\tau{({Q_\tau}^A)^2{\tilde{R}_J(\tau)}} \nonumber\\
 &  & + {\frac{Q_\lambda^A Q_\mu^A}{(\lambda
-\mu)}}\left(R_J(\lambda)-R_J(\mu)\right)
+ {\frac{Q_\mu^A Q_\nu^A}{(\mu
-\nu)}}\left(R_J(\mu)-R_J(\nu)\right)\nonumber\\
& & + {\frac{Q_\nu^A Q_\lambda^A}{(\nu
-\lambda)}}\left(R_J(\nu)-R_J(\lambda)\right)] 
\end{eqnarray}
\begin{eqnarray}
{\bf F^B}_{xx} & = & {\frac{-4x^2}{3\pi(1-m_0)m_0^2}}[
\sum_{\tau^\prime}{(Q_{\tau^\prime}^B)^2 \tilde{Q}_{\tau^\prime}^B
 R_J(\tau^\prime/m_0^2)} -{\frac 1 {m_0^2}}
\sum_{\tau^\prime}{(Q_{\tau^\prime}^B)^2\tilde{R}_J(\tau^\prime/m_0^2)}
 \nonumber \\
&  & +{\frac{Q_{\lambda^\prime}^B Q_{\mu^\prime}^B}{(\lambda^\prime
-\mu^\prime)}}
\left(R_J(\lambda^\prime/m_0^2)-R_J(\mu^\prime/m_0^2)\right)
+ {\frac{Q_{\mu^\prime}^B Q_{\nu^\prime}^B}{(\mu^\prime -\nu^\prime)}}
\left(R_J(\mu^\prime/m_0^2)-R_J(\nu^\prime/m_0^2)\right) \nonumber\\
& & + {\frac{Q_{\nu^\prime}^B Q_{\lambda^\prime}^B}{(\nu^\prime
-\lambda^\prime)}}
\left(R_J(\nu^\prime/m_0^2)-R_J(\lambda^\prime/m_0^2)\right)].
 \end{eqnarray}

The force derivatives ${\bf F}_{yy}$ and ${\bf F}_{zz}$ are obtained
by substituting ($S_{\tau}^A, S_{\tau^\prime}^A$) and
($T_{\tau}^A, T_{\tau^\prime}^B$) (\ref{qst}) in place of
($Q_{\tau}^A, Q_{\tau^\prime}^A$) respectively.
The cross derivatives of the forces may be written
\begin{eqnarray}
{\bf F}_{xy}  & = & {\bf F^A}_{xy} + {\bf F^B}_{xy}, 
\end{eqnarray}
\begin{eqnarray}
{\bf F^A}_{xy} & = & {\frac{4xy}{3\pi(1-m_0)}}[
\sum_\tau{(Q_\tau^A S_\tau^A) \tilde{Q}_\tau^A R_J(\tau)} -
\sum_\tau{(Q_\tau^A S_\tau^A)\tilde{R}_J(\tau)} \nonumber\\
& & + {\frac{Q_\lambda^A S_\mu^A}{(\lambda
-\mu)}}\left(R_J(\lambda)-R_J(\mu)\right)
+ {\frac{Q_\mu^A S_\nu^A}{(\mu
-\nu)}}\left(R_J(\mu)-R_J(\nu)\right)\nonumber\\
& & + {\frac{Q_\nu^A S_\lambda^A}{(\nu
-\lambda)}}\left(R_J(\nu)-R_J(\lambda)\right)] 
\end{eqnarray}
\begin{eqnarray}
{\bf F^B}_{xy} & = & {\frac{-4xy}{3\pi(1-m_0)m_0^2}}[
\sum_{\tau^\prime}{(Q_{\tau^\prime}^B S_{\tau^\prime}^B)
 \tilde{Q}_{\tau^\prime}^B R_J(\tau^\prime/m_0^2)} -{\frac 1 {m_0^2}}
\sum_{\tau^\prime}{(Q_{\tau^\prime}^B S_{\tau^\prime}^B)
\tilde{R}_J(\tau^\prime/m_0^2)} \nonumber \\
& & + {\frac{Q_{\lambda^\prime}^B S_{\mu^\prime}^B}{(\lambda^\prime
-\mu^\prime)}}
\left(R_J(\lambda^\prime/m_0^2)-R_J(\mu^\prime/m_0^2)\right)
+ {\frac{Q_{\mu^\prime}^B S_{\nu^\prime}^B}{(\mu^\prime -\nu^\prime)}}
\left(R_J(\mu^\prime/m_0^2)-R_J(\nu^\prime/m_0^2)\right) \nonumber \\
& &+ {\frac{Q_{\nu^\prime}^B S_{\lambda^\prime}^B}{(\nu^\prime
-\lambda^\prime)}}
\left(R_J(\nu^\prime/m_0^2)-R_J(\lambda^\prime/m_0^2)\right)].
\end{eqnarray}
Similar expressions may be obtained  for ${\bf F}_{yz}$ and ${\bf F}_{zx}$.

A high degree of speed and accuracy were desired for the orbit integration
routines. The accuracy with which the forces are computed and
consequently the speed of the orbital integration depends
 crucially on the accuracy with which the ellipsoidal coordinates
 ($\lambda, \mu, \nu$) are
calculated. 
The requisite accuracy could always be obtained using the NAG routine
C02AGF for finding the roots of a polynomial equation;
however these root evaluations
were computationally expensive. Therefore for large values of the
constant $m_0$ the roots were computed from standard algebraic
expressions for the roots of a cubic equation (e.g. Press et al.
1987). When $m_0$ is small the algebraic expressions
to evaluate the roots ($\lambda^\prime, \mu^\prime, \nu^\prime)$
 do not yield the required accuracy and the NAG routines are essential.

A great increase in the speed of evaluation of the forces was obtained
by fitting cubic splines (NAG routine E01BAF) to the elliptic 
integrals
 $R_J(\tau)$, 
$R_J(\tau^\prime/m_0^2)$ and their derivatives $\tilde{R}_J(\tau)$,
$\tilde{R}_J(\tau^\prime/m_0^2)$.
$\tau$ lies in the range $ c^2 \le \tau \le a^2 \le \lambda_{max}$ and
$m_0^2 c^2 \le \tau^\prime \le m_0^2a^2 \le \lambda^\prime_{max}$,
 where $\lambda_{max}$
and $\lambda^\prime_{max}$ are the maximum values of the ellipsoidal
coorinates $\lambda$ and $\lambda^\prime$ which depend on the
maximum values of the cartesian coordinates ($x^2, y^2, z^2$).
The integral are first tabulated over the range
$[c^2, \lambda_{max}/m_0^2]$  at the start of the orbit integration
and are evaluated by a cubic spline interpolant (NAG routine E02BBF)
thereafter. The functions $R_J(\tau)$, $R_J(\tau^\prime/m_0^2)$,
$\tilde{R}_J(\tau)$ and
$\tilde{R}_J(\tau^\prime/m_0^2)$ increase steeply as $\tau$ decreases for small values of
$\tau$ and it was found that the accuracy of interpolation could be greatly
improved by multiplying the functions by powers of $\tau$ and
 $\tau^\prime/m_0^2$
respectively. The actual functions that were splined were therefore:
[$(\tau)^{3/4}R_J(\tau)$], [$(\tau^\prime/m_0^2)^{3/4}R_J(\tau^\prime/m_0^2)$]
 and 
[$(\tau)^{3/2}\tilde{R}_J(\tau)$],
[ $(\tau^\prime/m_0^2)^{3/2}\tilde{R}_J(\tau^\prime/m_0^2)$].
This proceedure yielded a value of the true functions which was accurate
to at least 14 figures.

The potential
due to a  central black hole of mass $M_{BH}$ at a radius $R$ from
the center is $\Phi^{BH} = - G M_{BH}/R$. In cartesian coordinates the
forces at a point ($x_1, x_2, x_3$) (with $R^2 = {\sum_{i=1,3} x_i^2}$
 and $G=1$) are
given by
\begin{eqnarray}
F^{BH}_{x_i} & = & {{-M_{BH} x_i}\over R^3}
\end{eqnarray}
The corresponding force derivatives are
\begin{eqnarray}
F^{BH}_{x_i x_j} & = & {{- M_{BH}}\over {R^5}}(\delta_{ij}R^2 - {3 x_i x_j})
\end{eqnarray}

\clearpage

\figcaption[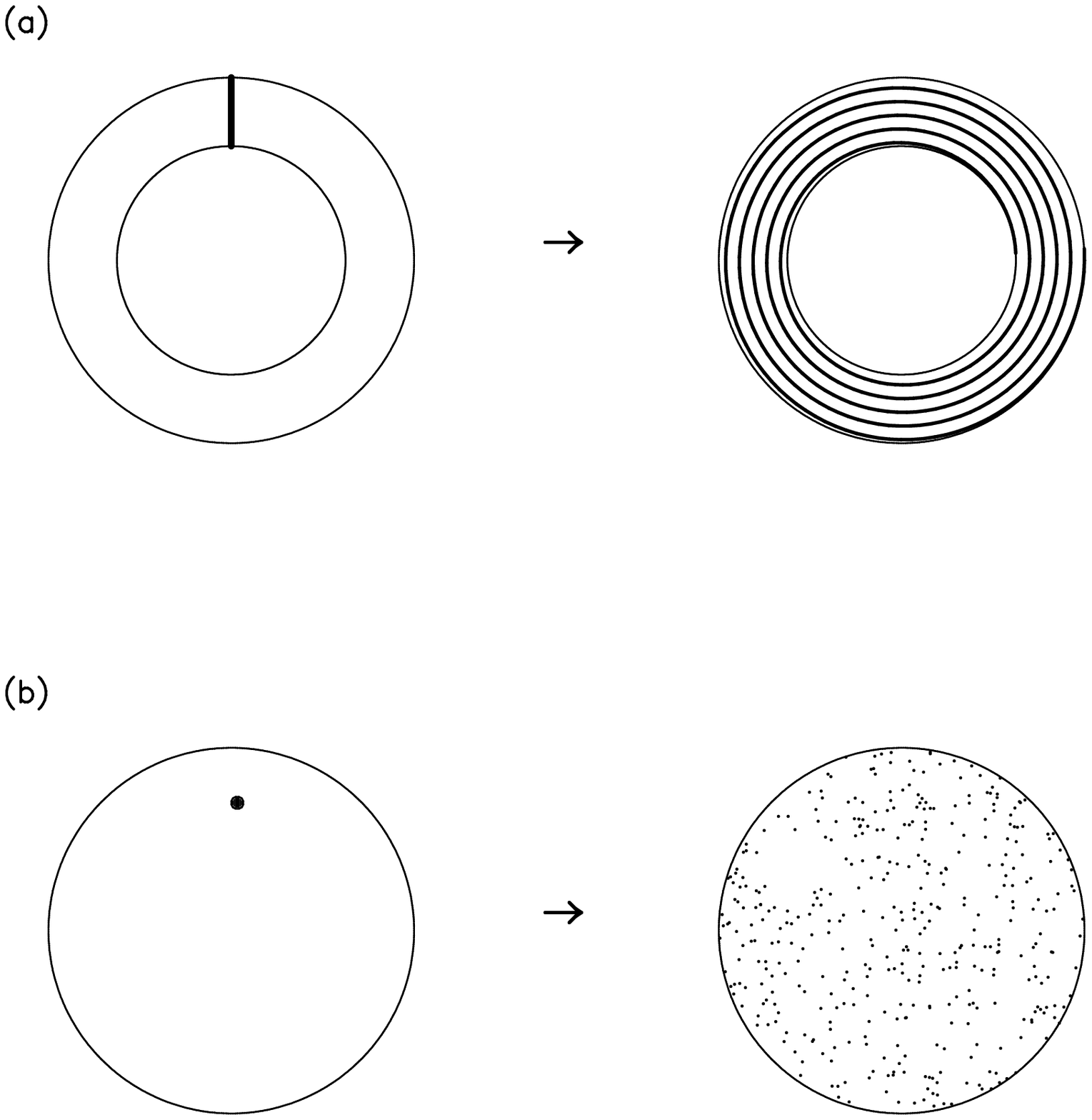]{\label{fig1}} Phase-mixing (a) vs. chaotic mixing (b).

\figcaption[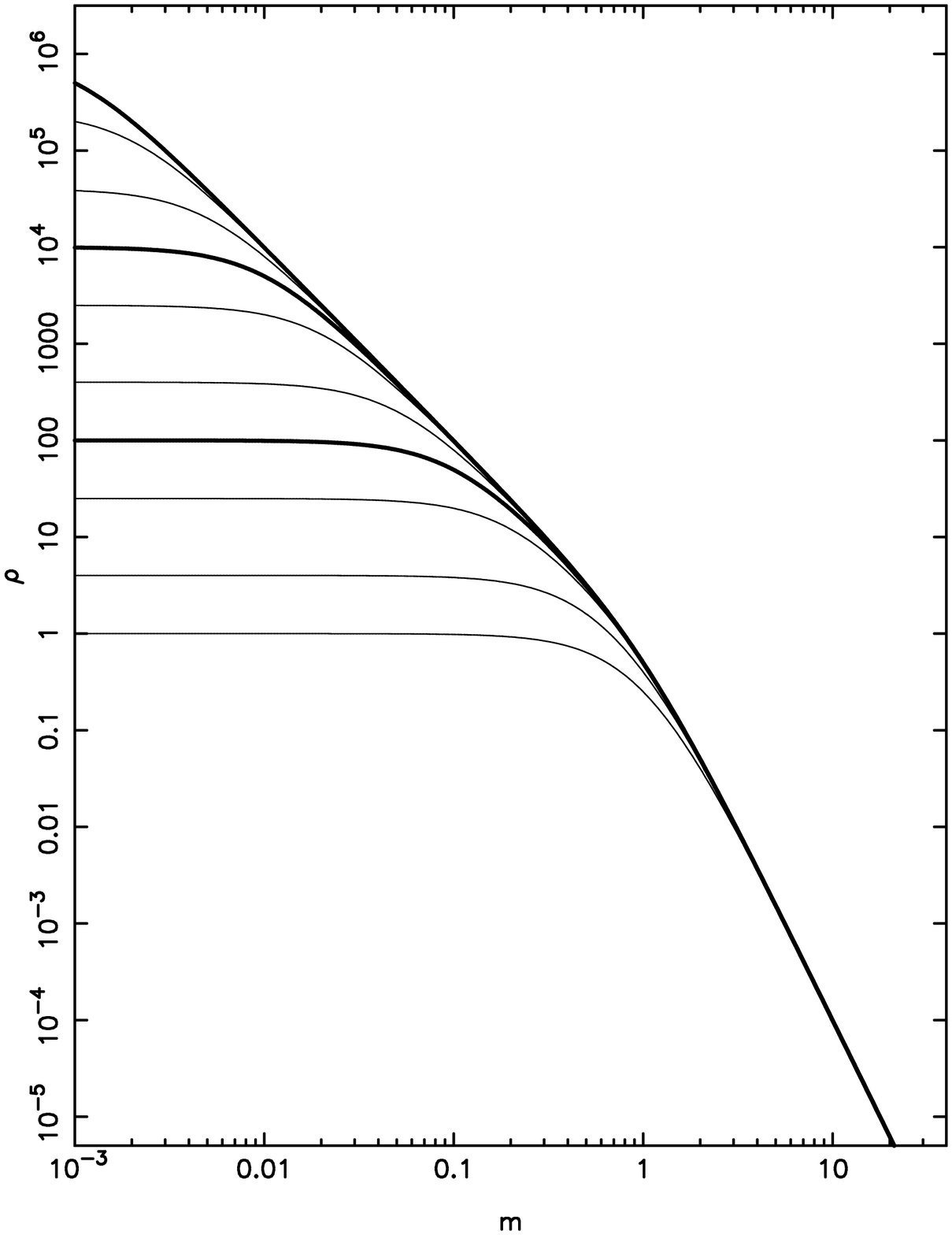]{\label{fig2}} Density law (5), for
$m_0=0,0.002,0.005,0.01,0.02,0.05,0.1,0.2,0.5,1$.
Heavy lines are profiles with the three values of $m_0$ used in
the potentials studied here.

\figcaption[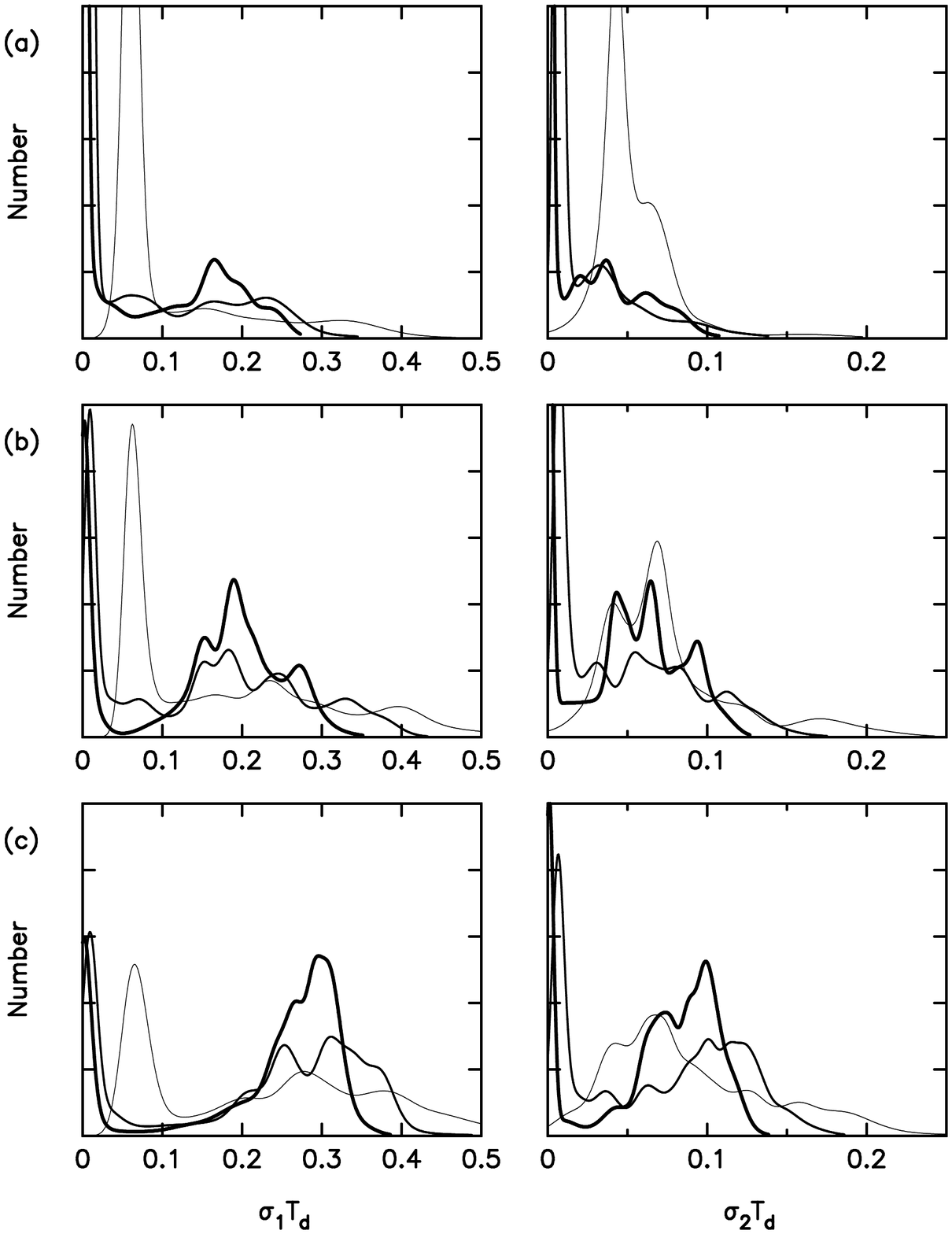]{\label{fig3}} 
Histograms of Liapunov numbers 
for isoenergetic ensembles of boxlike orbits in the triaxial 
potential of Eq. (8), with $m_0=10^{-1}$ (a), 
$m_0=10^{-2}$ (b) and $m_0=10^{-3}$ (c).
$c/a=0.5$ and $b/a=0.79$.
Heavy lines: $t=10^4 T_D$; intermediate lines: $t=10^3 T_D$; thin 
lines: $t=10^2T_D$.

\figcaption[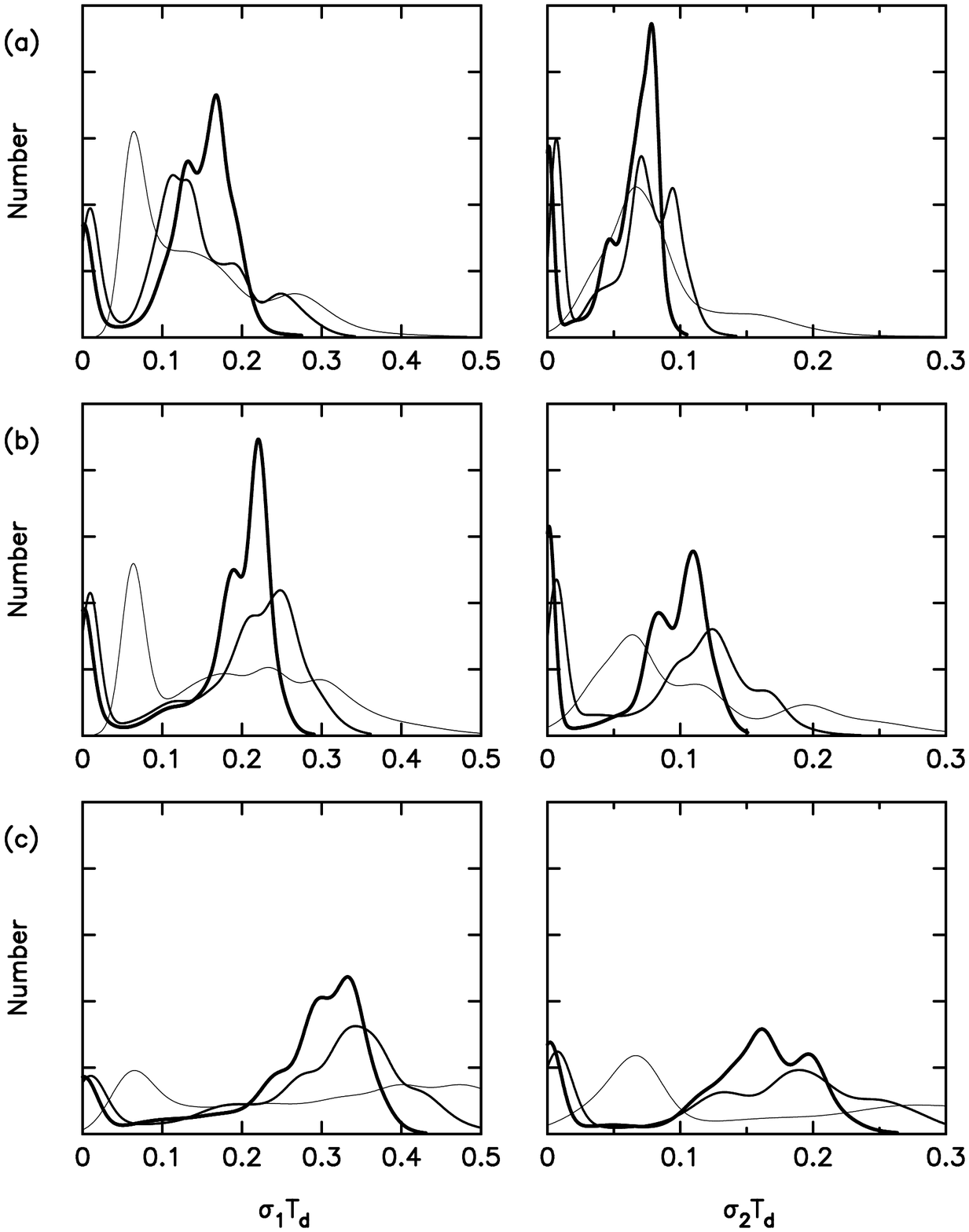]{\label{fig4}} 
Histograms of Liapunov numbers 
for isoenergetic ensembles of boxlike orbits in the triaxial 
potential of Eq. (8), with $m_0=10^{-1}$ and
$M_{BH}=10^{-3}$ (a), $3\times 10^{-3}$ (b) and $10^{-2}$ (c).
$c/a=0.5$ and $b/a=0.79$.
Heavy lines: $t=10^4 T_D$; intermediate lines: $t=10^3 T_D$; thin 
lines: $t=10^2T_D$.

\figcaption[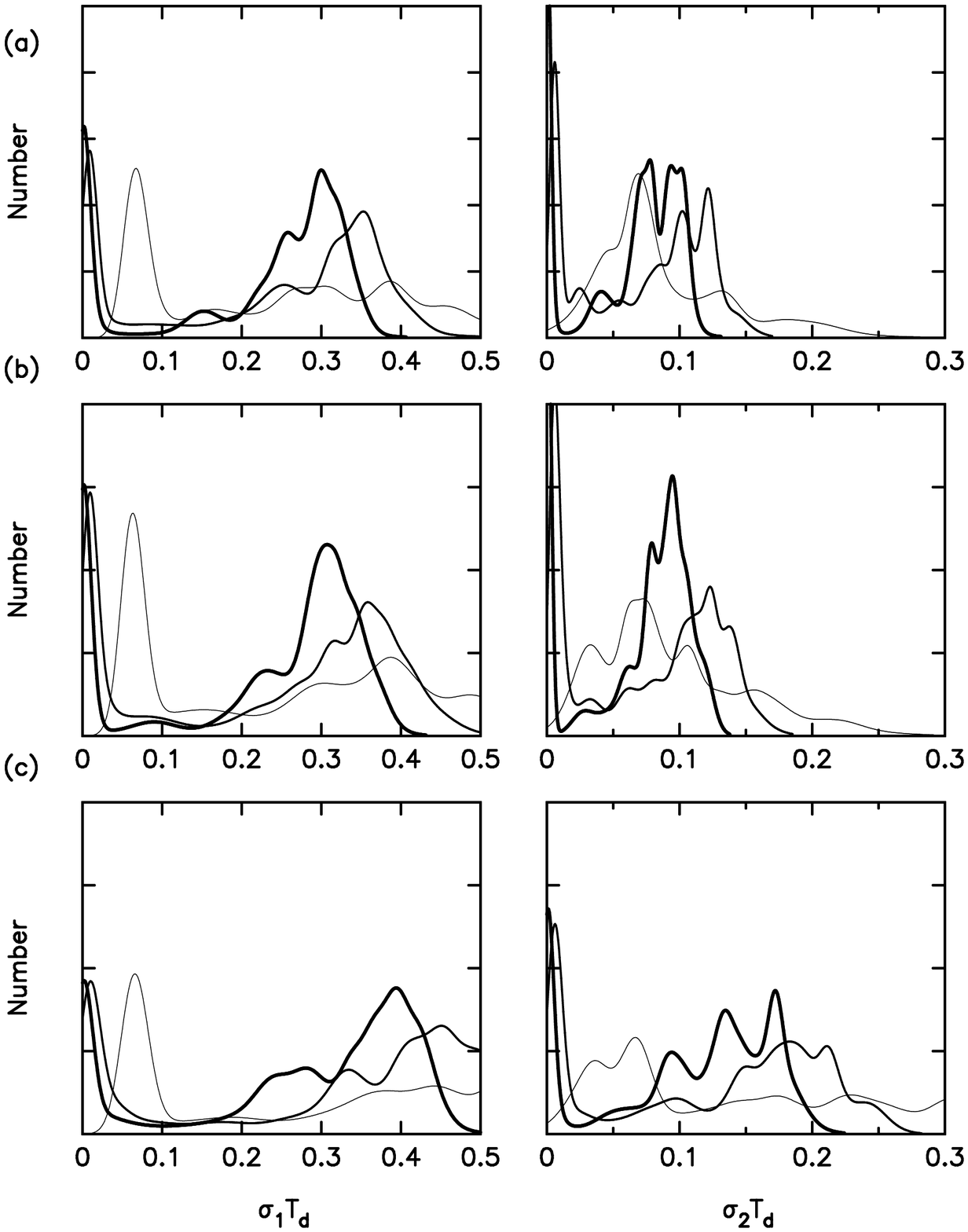]{\label{fig5}}
Histograms of Liapunov numbers 
for isoenergetic ensembles of boxlike orbits in the triaxial 
potential of Eq. (8), with $m_0=10^{-3}$ and
$M_{BH}=10^{-3}$ (a), $3\times 10^{-3}$ (b) and $10^{-2}$ (c).
$c/a=0.5$ and $b/a=0.79$.
Heavy lines: $t=10^4 T_D$; intermediate lines: $t=10^3 T_D$; thin 
lines: $t=10^2T_D$.

\figcaption[figure6.ps]{\label{fig6}}
Velocities at central crossings $(x\approx y\approx z\approx 0)$ 
for boxlike orbits in three models.
The dots in this velocity octant map mark the velocity at one near-center
passage.
(a) $M_{BH}=0$.
Large dots: $m_0=10^{-1}$; medium dots: $m_0=10^{-2}$; small dots: 
$m_0=10^{-3}$.
(b) $m_0=10^{-1}$.
Large dots: $M_{BH}=10^{-3}$; medium dots: $M_{BH}=3\times 10^{-3}$;
small dots: $M_{BH}=10^{-2}$.

\figcaption[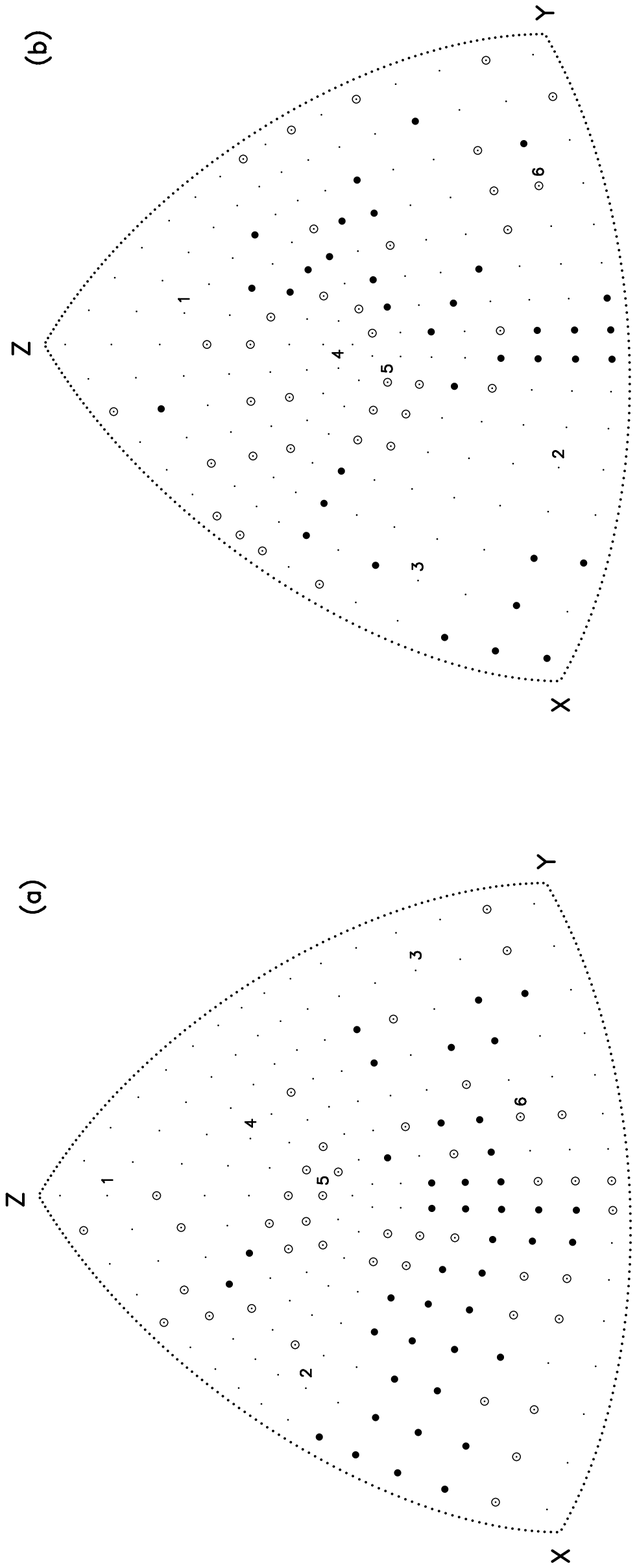]{\label{fig7}}
Starting points of boxlike orbits on one octant of the equipotential
surface.
Small dots are stochastic orbits; large dots are regular orbits; circles
are trapped stochastic orbits.
(a) $m_0=10^{-3}$, $M_{BH}=0$ (Model 1).
(b) $m_0=10^{-1}$, $M_{BH}=3\times 10^{-3}$ (Model 2).
Numbers denote starting points of the ensembles used in the mixing
experiments of \S 5.

\figcaption[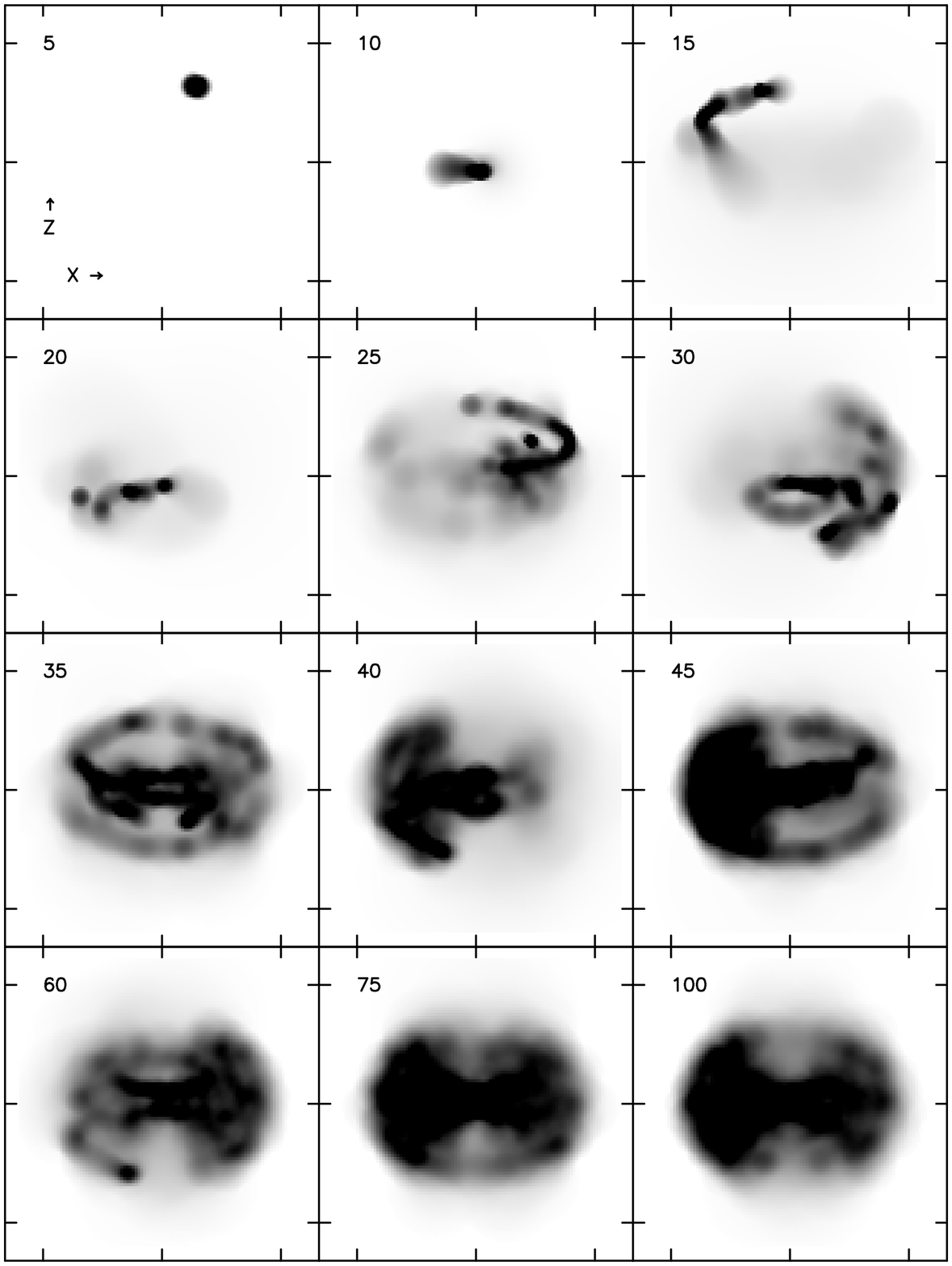]{\label{fig8}}
Evolution of Ensemble No. 4 in Model 1 ($m_0=10^{-3}$, $M_{BH}=0$).
Plotted are projections onto the $x-z$ plane of the configuration-space 
density of $10^4$, independently-moving points.
Numbers are the elapsed time in units of $T_D$.
Tick marks are separated by one length unit.

\figcaption[figure9.ps]{\label{fig9}}
Cuts through the principal planes of the ensemble densities at $t=200T_D$.
(a) Ensemble No. 1, Model 1. (b) Ensemble No. 1, Model 2.

\figcaption[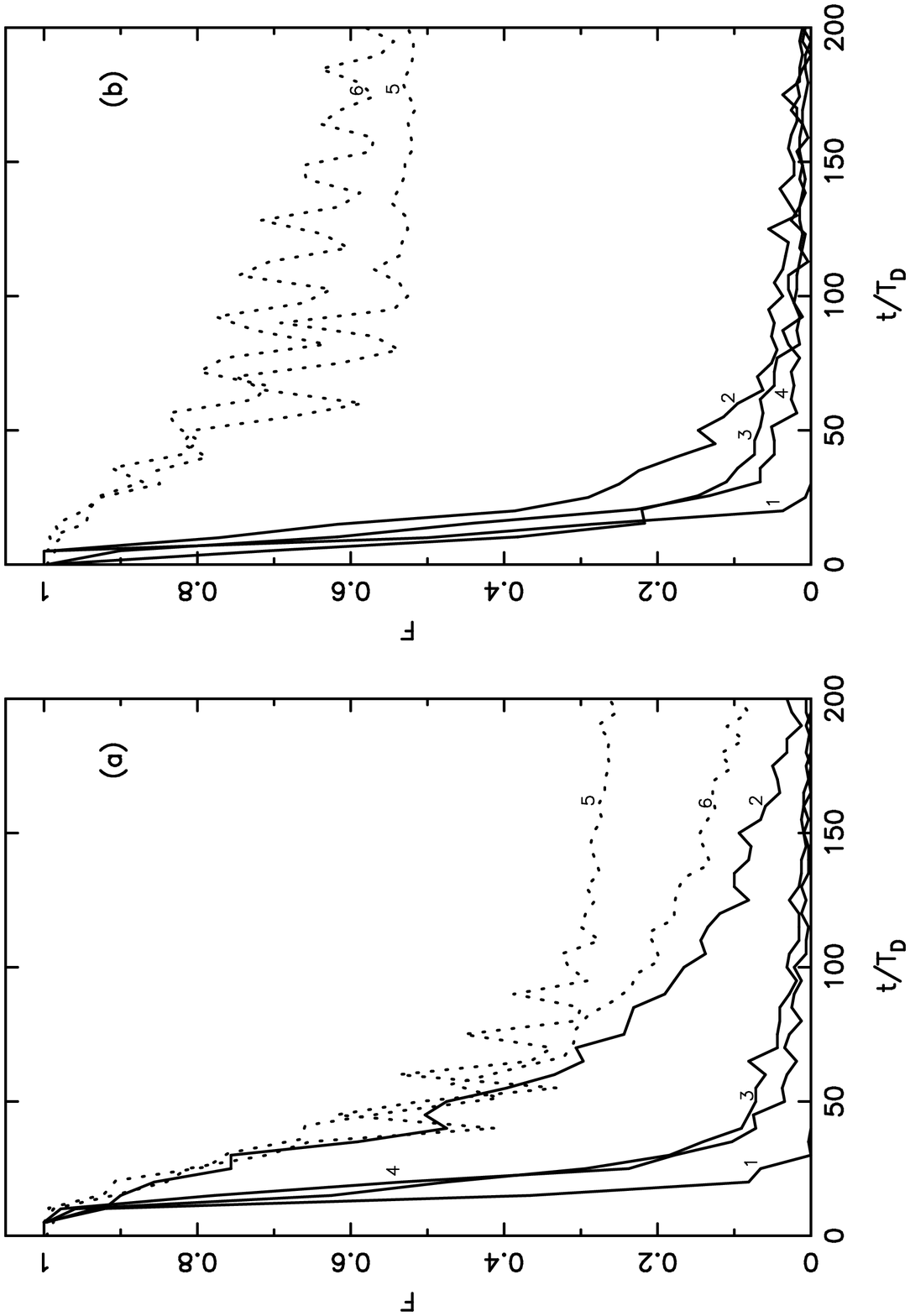]{\label{fig10}}
Evolution of $F$, the fraction of configuration-space cells containing
no particles, for the 12 mixing ensembles.
(a) Model 1 ($m_0=10^{-3}$, $M_{BH}=0$); (b) Model 2 
($m_0=10^{-1}$, $M_{BH}=3\times 10^{-3}$).
Dotted lines are trapped ensembles.

\figcaption[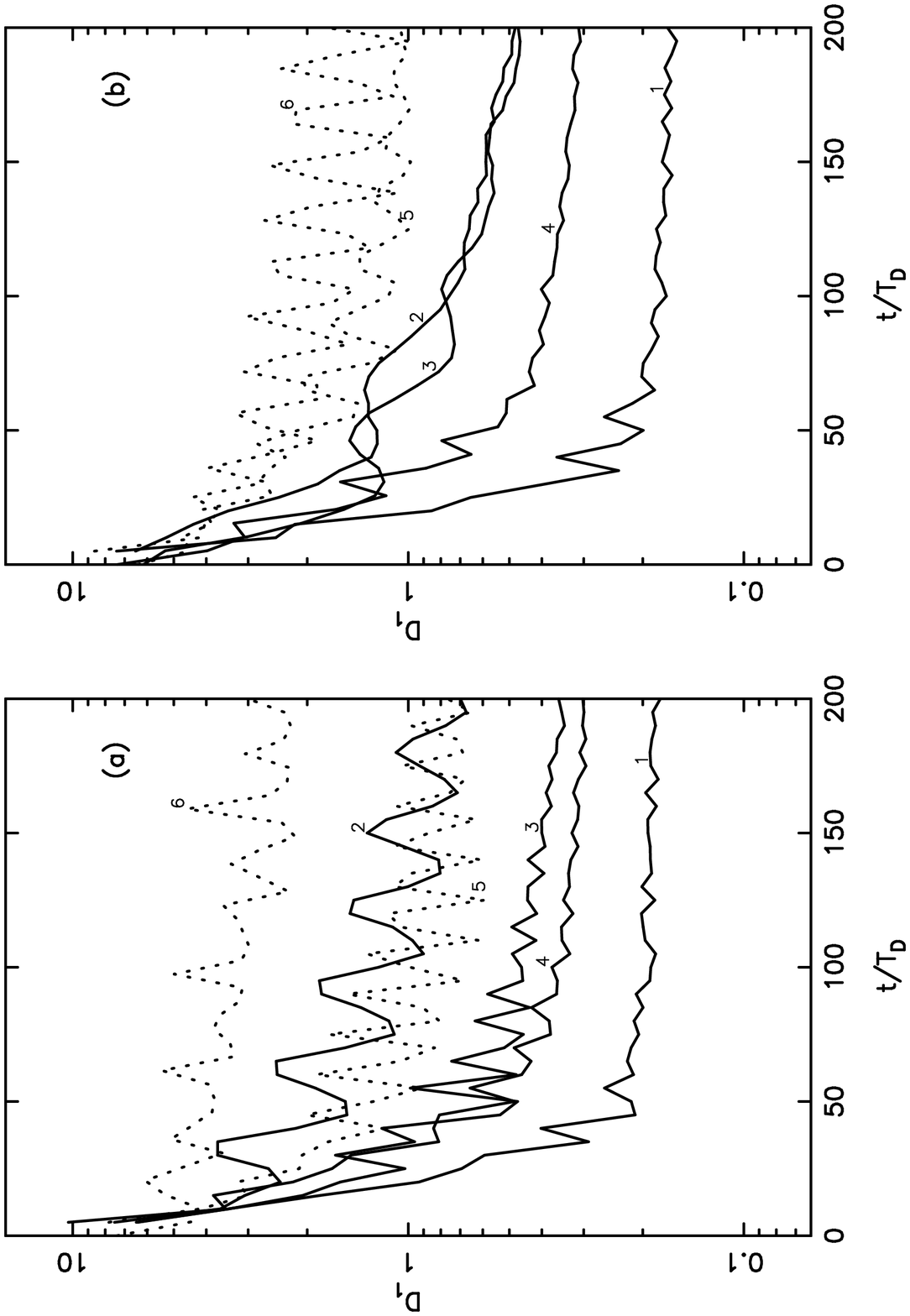]{\label{fig11}}
Evolution of $D_1$, the ``distance'' between the ensemble density and
the micro-canonical density, for the 12 mixing ensembles without noise.
(a) Model 1 ($m_0=10^{-3}$, $M_{BH}=0$); (b) Model 2 
($m_0=10^{-1}$, $M_{BH}=3\times 10^{-3}$).
Dotted lines are trapped ensembles.

\figcaption[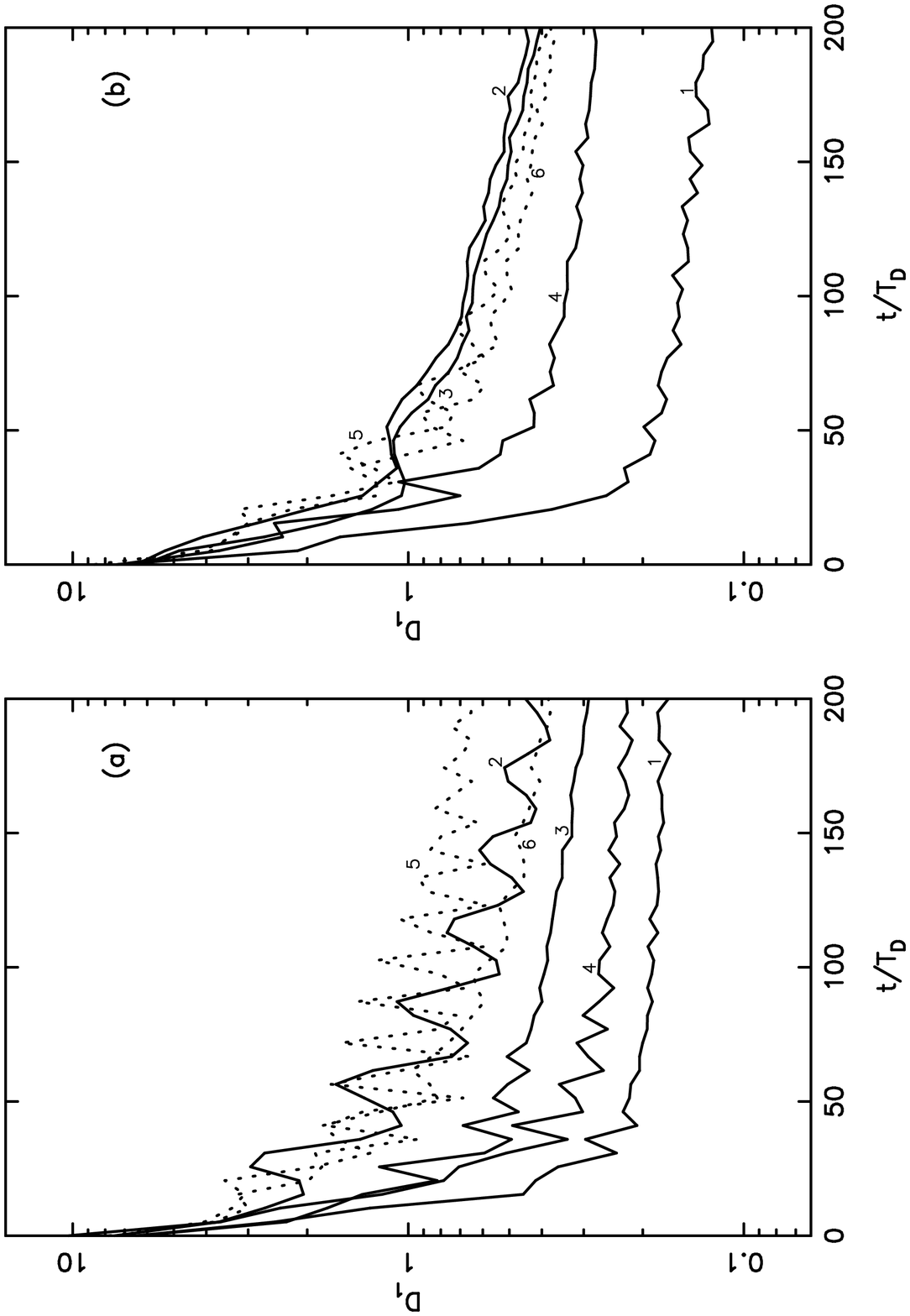]{\label{fig12}}
Evolution of $D_1$, the ``distance'' between the ensemble density and
the micro-canonical density, for the 12 mixing ensembles with noise,
$\eta=10^4$.
(a) Model 1 ($m_0=10^{-3}$, $M_{BH}=0$); (b) Model 2 
($m_0=10^{-1}$, $M_{BH}=3\times 10^{-3}$).
Dotted lines are trapped ensembles.

\setcounter{figure}{0}

\begin{figure}
\plotone{figure1.ps}
\caption{ }
\end{figure}

\begin{figure}
\plotone{figure2.ps}
\caption{ }
\end{figure}

\begin{figure}
\plotone{figure3.ps}
\caption{ }
\end{figure}

\begin{figure}
\plotone{figure4.ps}
\caption{ }
\end{figure}

\begin{figure}
\plotone{figure5.ps}
\caption{ }
\end{figure}

\begin{figure}
\plotone{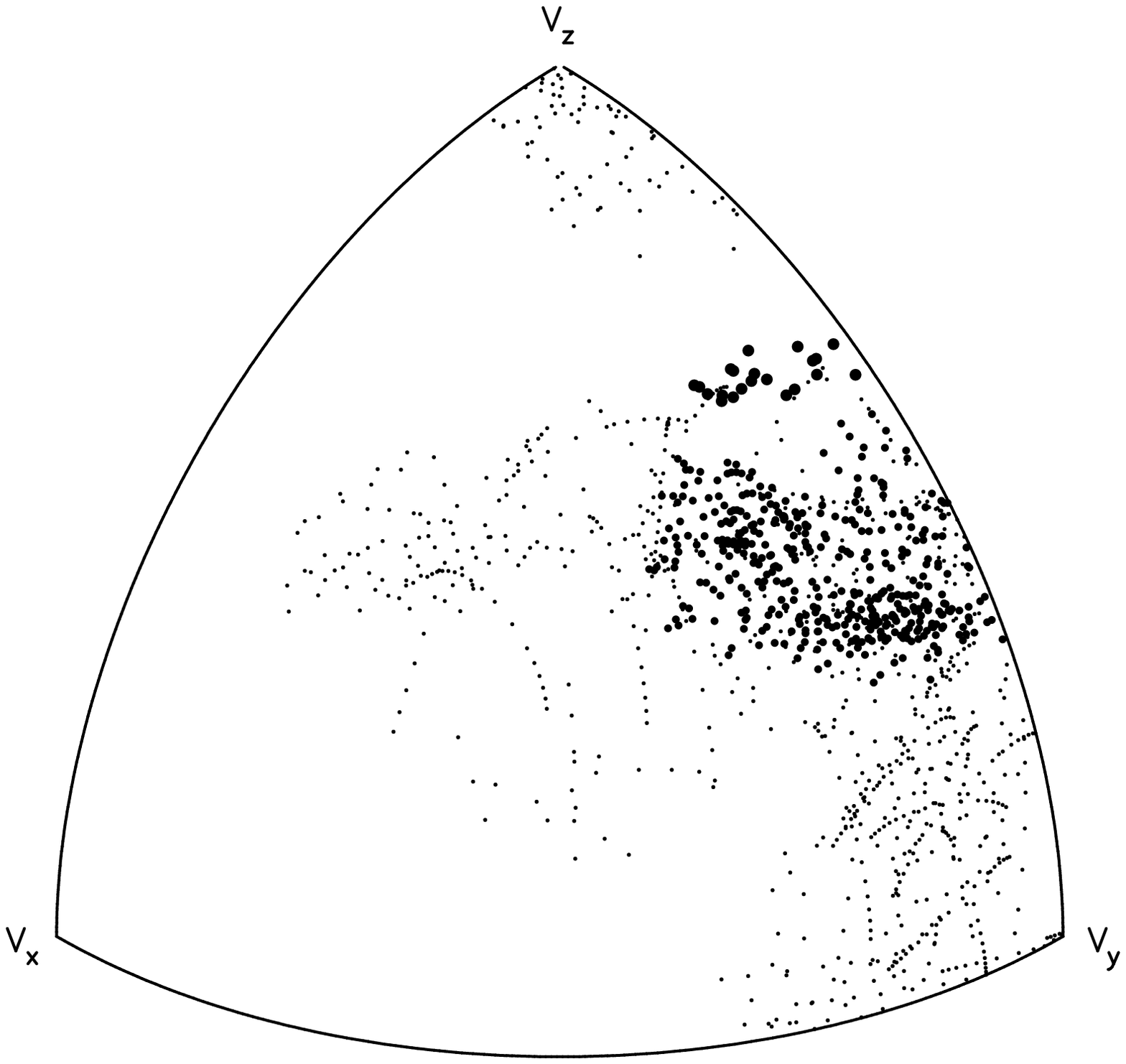}
\caption{ }
\end{figure}

\setcounter{figure}{5}
\begin{figure}
\plotone{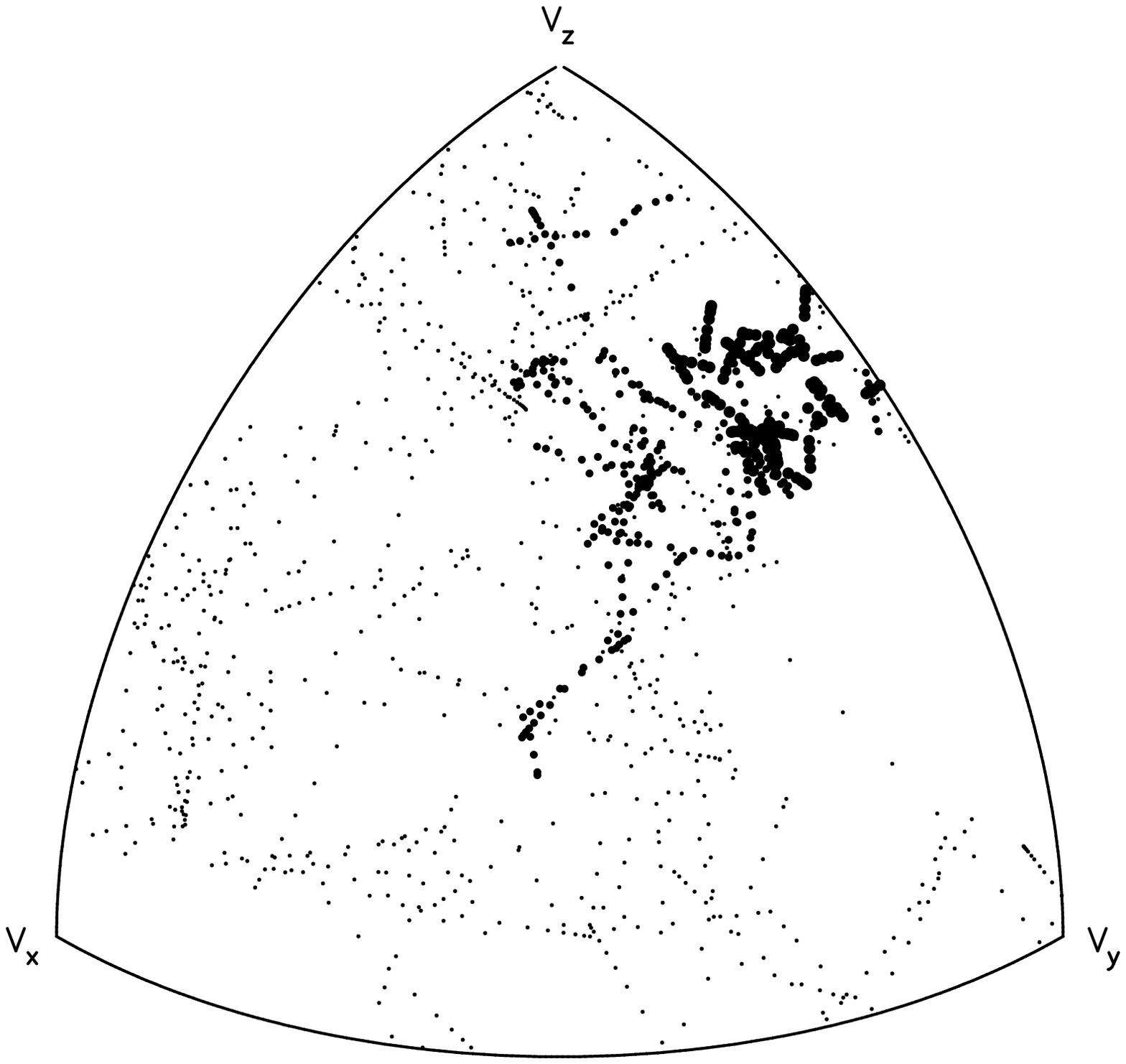}
\caption{ }
\end{figure}

\begin{figure}
\plotone{figure7.ps}
\caption{ }
\end{figure}

\begin{figure}
\plotone{figure8.ps}
\caption{ }
\end{figure}

\begin{figure}
\plotone{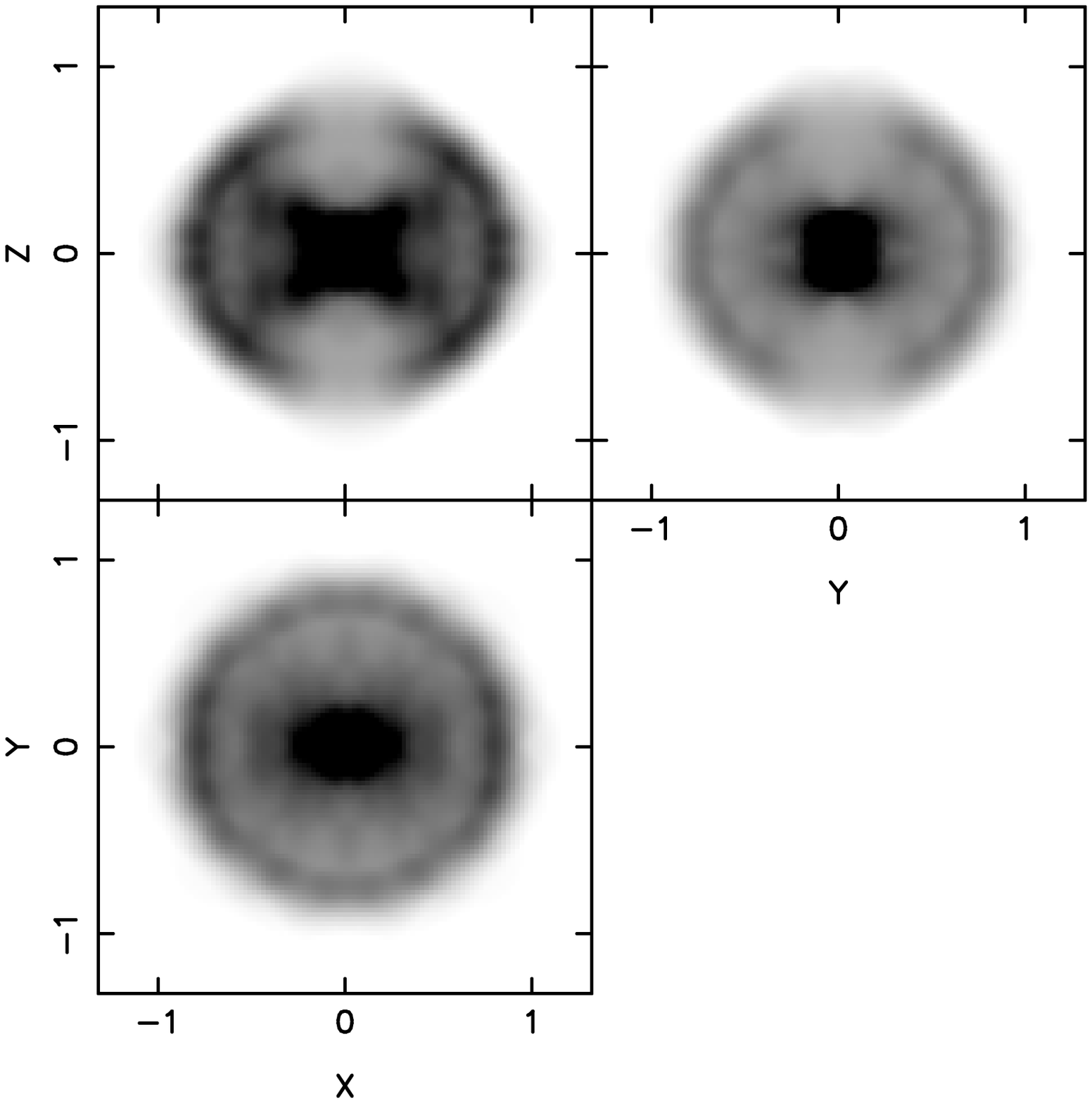}
\caption{ }
\end{figure}

\setcounter{figure}{8}
\begin{figure}
\plotone{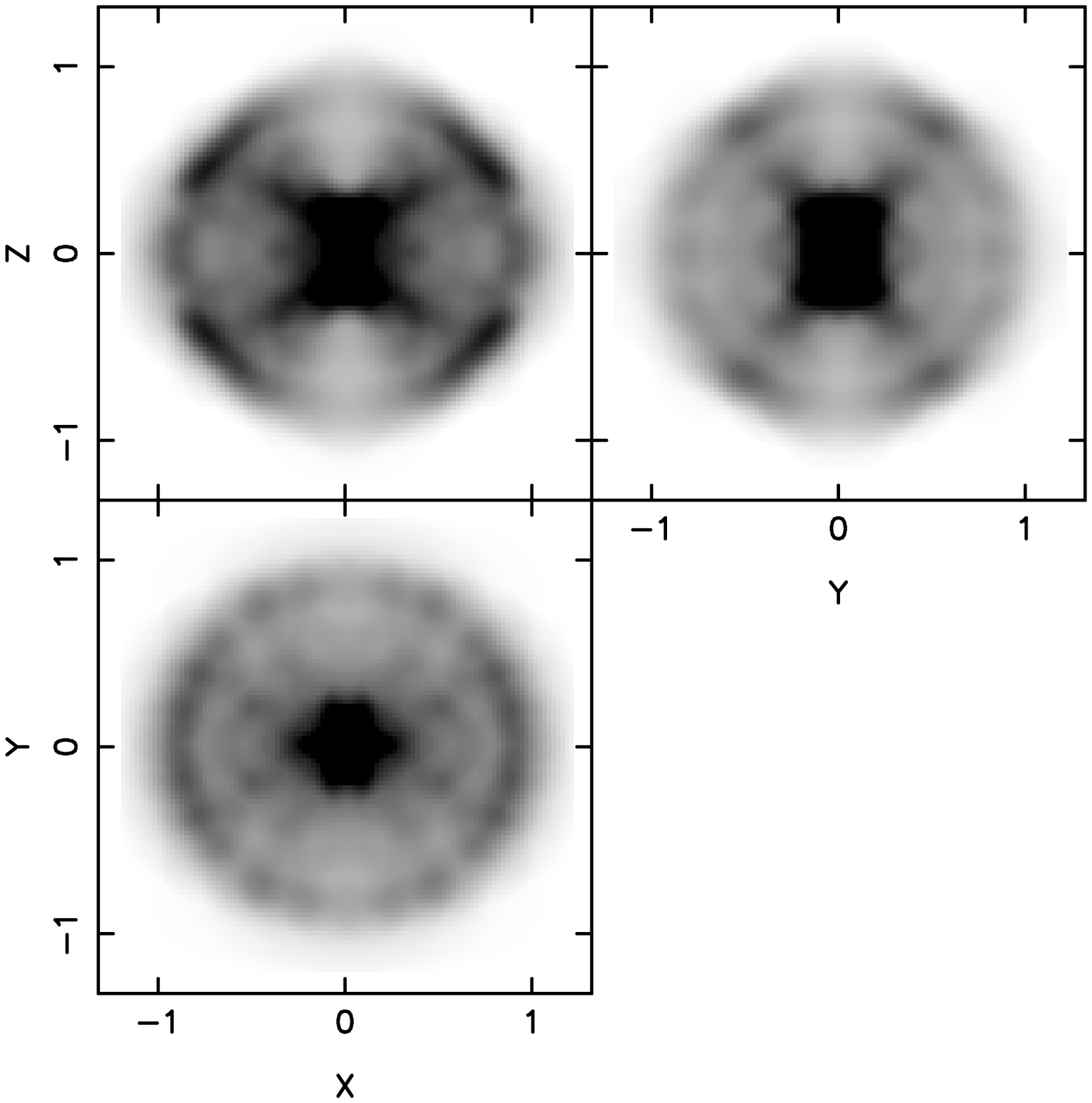}
\caption{ }
\end{figure}

\begin{figure}
\plotone{figure10.ps}
\caption{ }
\end{figure}

\begin{figure}
\plotone{figure11.ps}
\caption{ }
\end{figure}

\begin{figure}
\plotone{figure12.ps}
\caption{ }
\end{figure}
 

\clearpage

\begin{references}

\reference{} Arnold, V. I. 1964, Russian Math. Surveys, 18, 85
\reference{} Arnold, V. I. 1989, Mathematical Methods of 
	Classical Mechanics (New York: Springer)
\reference{} Arnold, V. I. \& Avez, A. 1968, Ergodic Problems of 
	Classical Mechanics (New York: Benjamin) 
\reference{} Benettin, G., Galgani, L, Giorgilli, A. \& Strelcyn, J.-M. 
	1980, Meccanica, 15, 21
\reference{} Binney, J. J. 1982a, MNRAS, 201, 1
\reference{} Binney, J. J. 1982b, MNRAS, 201, 15
\reference{} Binney, J. J. \& Tremaine, S. 1987, Galactic Dynamics 
	(Princeton University Press), 220
\reference{} Chandrasekhar, S. 1969, Ellipsoidal Figures of Equilibrium (New 
York: Dover), 52
\reference{} Crane, P., \etal~ 1993, AJ, 106, 1371
\reference{} Davies, R. \etal~ 1983, ApJ, 266, 41
\reference{} Dehnen, W. 1993, MNRAS, 265, 250
\reference{} Dehnen, W. 1995, MNRAS, 274, 919
\reference{} Dehnen, W. 1996, private communication
\reference{} Dejonghe, H. 1987, ApJ, 320, 477
\reference{} de Zeeuw, P. T. 1985, MNRAS, 216, 273
\reference{} de Zeeuw, P. T. 1994, Lecture Notes of the 
	Canary Islands Winter School, ``Formation of Galaxies.''
\reference{} de Zeeuw, P. T. \& Lynden-Bell, D. 1985, MNRAS, 215, 713
\reference{} de Zeeuw, P. T. \& Pfenniger, D. 1988, MNRAS, 235, 949
\reference{} Ferrarese, L, van den Bosch, F. C., Ford, H. C., Jaffe, W., \& 
	O'Connell, R. W. 1994, AJ, 108, 1598
\reference{} Ford, H. C. \etal~ 1994, ApJL 435, L27
\reference{} Franx, M., Illingworth, G. \& de Zeeuw, T. 1991, 
	ApJ, 383, 112
\reference{} Gebhardt, K. \etal~ 1996, preprint
\reference{} Gerhard, O. E. 1993, 6th European EADN 
	Summer School: Galactic Dynamics and N-Body Simulations, ed. G. 
	Contopoulos \& G. Spyrou (New York: Springer)
\reference{} Gerhard, O. E. \& Binney, J. J. 1985, MNRAS, 216, 
	467
\reference{} Goodman, J. \& Schwarzschild, M. 1981, ApJ, 245, 1087
\reference{} Habib, S., Kandrup, H. E., \& Mahon, M. E. 1995, 
	Phys. Rev. E, in press
\reference{} Hubble, E. W. 1930, ApJ, 71, 231
%\reference{} Jaffe, W., Ford, H. C., O'Connell, R. W. van den Bosch, F. C. \& 
%	Ferrarese, L. 1994, AJ, 108, 1567
\reference{} Jeans, J. J. 1915, MNRAS, 76, 71
\reference{} Kandrup, H. E. \& Mahon, M. E. 1994, Phys. Rev. E, 
	49, 3735 
\reference{} Karney, C. F. F. 1983, Physica 8D, 360
\reference{} Katz, N. \& Richstone, D. O. 1985, ApJ, 296, 331
\reference{} Kormendy, J., \etal~ 1995, ESO/OHP Workshop on Dwarf Galaxies, 
	ed. G. Meylan \& P. Prugniel (Garching: ESO), 147
\reference{} Kormendy, J. \& Bender, R. 1996, preprint
\reference{} Kormendy, J. \& Richstone, D. 1995, Ann. Rev. 
	Astron. Astrophys., 33, 581
\reference{} Krylov, N. S. 1979, Works on the Foundations of 
	Statistical Physics (Princeton University Press)
\reference{}Kuzmin, G. G. 1973, The Dynamics of Galaxies and Star Clusters, 
	ed. T. B. Omarov (Nauka of the Kazakh S. S. R., Alma-Ata), 71.
\reference{} Lauer, T., \etal~ 1995, AJ, 110, 2622 
\reference{} Lichtenberg, A. J. \& Lieberman, M. A. 1983, Regular and Stochastic
	Motion (New York: Springer)
\reference{} Lynden-Bell, D. 1967, MNRAS, 136, 101
\reference{} Mackay, R. S, Meiss, J. D. \& Percival, I. C. 1984, Physica 13D, 55
\reference{} Mahon, M. E., Abernathy, R. A., Bradley, B. O. \& 
	Kandrup, H. E. 1995, MNRAS, 275, 443
\reference{} Merritt, D. 1980, ApJ Suppl., 43, 435
\reference{} Merritt, D. \& Fridman, T. 1995, A. S. P. Conf. 
	Ser. Vol. 86, Fresh Views of Elliptical Galaxies, ed. A. Buzzoni, 
	A. Renzini \& A. Serrano (Provo: Astronomical Society of the 
	Pacific), 13
\reference{} Merritt, D. \& Fridman, T. 1996, ApJ, 460, 136
\reference{} Miralda-Escude, J. \& Schwarzschild, M. 1989, ApJ, 
	339, 752
\reference{} Miyoshi, M., Moran, J., Herrnstein, J., Greenhill, L.,
	Nakai, N., Diamond, P. \& Inoue, M. 1995, Nature, 373, 127
\reference{} Moller, P., Stiavelli, M. \& Zeilinger, W. W. 1995, 
	MNRAS, 276, 979
\reference{} Ott, E. 1993, Chaos in Dynamical Systems (Canbridge 
	University Press), 257
\reference{} Schwarzschild, M. 1981, The Structure and 
	Evolution of Normal Galaxies, ed. S. M. Fall \& D. Lynden-Bell 
	(Cambridge University Press), 43
\reference{} Schwarzschild, M. 1987, Ann. N. Y. Acad. Sci., 497, 16
\reference{} Schwarzschild, M. 1993, ApJ, 409, 563
\reference{} Sinai, Ya. G. 1976, Introduction to Ergodic Theory 
	(Princeton University Press)
\reference{} Spitzer, L. \& Hart, M. 1971, ApJ, 164, 399
\reference{} Tonry, J. L. 1987, ApJ, 322, 632
\reference{} Tremblay, B. \& Merritt, D. 1996, AJ, 111, 000
\reference{} Udry, S. \& Pfenniger, D. 1988, AAp, 198, 135
\reference{} van der Marel, R. P., Evans, N. W., Rix, H. W., White,
	S. D. M. \& de Zeeuw, T. 1994, MNRAS, 271, 99

\end{references}
\end{document}